\documentclass[aps, 10pt, prd, twocolumn, showpacs,longbibliography]{revtex4-2}
\usepackage{amsmath}
\usepackage{amssymb}
\usepackage{graphicx}
\usepackage{bm}
\usepackage{lmodern}
\usepackage{verbatim}
\usepackage{units}
\usepackage{graphicx}
\usepackage[utf8]{inputenc}
\usepackage[svgnames]{xcolor}
\usepackage{soul}

\usepackage[english]{babel} 

\DeclareMathOperator\erf{erf}

\renewcommand{\vec}{\boldsymbol}

\definecolor{darkgreen}{rgb}{0.0, 0.5, 0.0}

\begin{document}

\title{A viable relativistic scalar theory of gravitation} 
\author{Diogo P. L. Bragança}
\email{braganca@stanford.edu}
\affiliation{Kavli Institute for Particle Astrophysics and Cosmology - KIPAC, Department of Physics, Stanford University, Stanford, CA 94305, USA}

\begin{abstract}
We build a self-consistent relativistic scalar theory of gravitation on a flat Minkowski spacetime from a general field Lagrangian. 
It is shown that, for parameters that satisfy the Equivalence Principle, this theory predicts the same outcome as general relativity for every classical solar-system test. 
This theory also admits gravitational waves that propagate at the speed of light, and the gravitational radiation energy loss in a binary system is shown to be very similar to the GR prediction.
We then analyze the strong gravity regime of the theory for a spherically symmetric configuration and find that there is an effective ``singularity'' near the Schwarzschild radius. 
The main goal of this work is to show that, contrary to what is commonly believed, there are relativistic scalar theories of gravitation defined on a Minkowski spacetime that are not ruled out by the classical solar system tests of general relativity.  
\end{abstract}

\maketitle

\tableofcontents

\section{Introduction}

The first attempts to build a relativistic theory of gravitation generalized Newton's theory to make it Lorentz invariant. 
The simplest generalization was to describe gravity as a scalar field $\Phi$, the gravitational potential, defined on a background Minkowski spacetime. 
The first successful consistent theories of this kind were Nordstr\"{o}m's two theories  \cite{Nordstrom1912,Nordstrom1913}, see also \cite{Wellner1964,Norton1992}. 
Consistency was automatically achieved because they could be derived from a Lagrangian. 
Even if this theory was defined in a Minkowski flat spacetime, it was also shown by Einstein and Fokker \cite{Einstein1914} to have a geometrical interpretation, since the particle's trajectories could be described as geodesics of a curved, conformally flat metric. 
The theory even satisfied the strong equivalence principle~\cite{Deruelle2011}.
However, it predicted a wrong precession of Mercury's perihelion and did not predict any light deflection, and so was discarded.
As general relativity appeared soon after and explained all solar system tests \cite{Einstein1916} (Mercury's perihelion precession, light bending, gravitational time dilation, and the Shapiro effect), see also \cite{Ni2016}, efforts to develop consistent and viable relativistic scalar theories dramatically decreased.

Nevertheless, scalar theories of gravitation where the metric is sourced by a scalar field have been developed even after general relativity either as useful tools to approximate general relativity \cite{Shapiro1993,Watt1999} or as legitimate theories on their own \cite{Bergmann1956,Pagetupper1968,Ni1973,Novello2012} (a compendium is given in~\cite{Ni1972}).
These theories are broadly divided in either stratified or Lorentz-invariant scalar theories~\cite{Ni1972}. 
The former break Lorentz invariance and have preferred frames, and the latter cannot explain solar system tests, in particular light bending.
The reason for that is the following (as explained in~\cite{Braganca2018}): in a scalar theory the interaction term between matter and gravity in the Lagrangian ${\cal L}_{\rm int}$ has to be ${\cal L}_{\rm int} \propto \Phi \, T$, where $T$ is the trace of the energy-momentum tensor, and this vanishes for light.
For this reason, in is often argued in the literature (see for example \cite{Bergmann1956,Gupta1957,Harvey1965,gravitation1973,Giulini2008}) that relativistic scalar theories of gravitation with no preferred frame cannot correctly describe the gravitational field.
In \cite{Novello2012}, the authors develop a theory with two metrics, where one of them depends on the gravitational field, and they explain the solar system tests.
Having said this, we are not aware of any Lorentz invariant scalar gravity theory (including among weak field limits of scalar-tensor theories) that treats the gravitational field exclusively in a Minkowski spacetime, and where the physical metric is purely emergent from this gravitational field in a natural way.

One may think that if a scalar theory of gravity had a Lorentz invariant Lagrangian that would allow for an interaction between the field and the individual components of $T^{ab}$, it might predict a light bending, and the above reason for discarding scalar theories would not be valid anymore. 
Indeed, such a theory could in principle also correctly explain the classical solar system tests of general relativity (GR).
That would definitely not mean that GR is incorrect. 
In fact, with gravitational wave astronomy, it has been proven right even in the strong gravity regime. 
Instead, it would just mean that a relativistic scalar field theory of gravitation defined on a Minkowski spacetime could predict the same outcomes for solar system tests as general relativity, or equivalently, that it could have the same Parametrized Post Newtonian (PPN) parameters used in solar system tests as general relativity. 
This PPN formalism is a standard method to compare relativistic theories of gravitation. 
It is defined in \cite{gravitation1973,Will1993} and is used for example in \cite{Will1972,Ni1972,Will1976,Will2014,Hohmann2013,Hohmann2016,Hohmann2017,Ni1973,Ni2016-2,Sanghai2017,Novello2012}. 
It would still be of theoretical interest to analyze a possibly viable scalar theory in the strong field regime and in gravitational wave emission, to see how it would compare to GR, and provide solid, definitive arguments as to why general relativity is a more accurate depiction of the gravitational field.

In this paper, we consider a relativistic scalar field theory of gravitation on a flat Minkowski spacetime with a general interaction Lagrangian that is the sum of a term proportional to $\Phi \, T$ and another term proportional to $ T^{ab} (\partial_a \Phi)(\partial_b \Phi)$. 
We consistently account for the self-interaction of the field (that is, gravitational energy gravitates) following an approach similar to that in \cite{Giulini1996,Franklin2014,Franklin2016}, and we enforce a version of the Equivalence Principle used by Nordstr\"{o}m~\cite{Nordstrom1913,Deruelle2011}.
We find that it predicts the correct results for every solar system test, a very similar gravitational wave emission from a system like the Hulse-Taylor binary pulsar, and that it could be different from general relativity in the strong field regime.

In Sec.~\ref{sec:lag}, we analyze the theory's Lagrangian and derive a condition from self-interaction using the field equation. 
In Sec.~\ref{sec:weak}, we give the weak field spherically symmetric vacuum solution. 
In Sec.~\ref{sec:partlag}, we obtain the particle Lagrangian and derive the equations of motion for a test particle. 
In Sec.~\ref{sec:ep}, the Equivalence Principle is used to impose more constraints on the theory, giving it predictive power in the post-Newtonian regime. 
In Sec.~\ref{sec:ppn}, we calculate the PPN parameters of the theory, by interpreting the trajectories of particles as geodesics of an effective metric. 
In Sec.~\ref{sec:light}, we provide a framework to study the propagation of light in this theory and compare it with GR.
In Sec.~\ref{sec:gw}, a calculation of the energy lost by gravitational wave emission by a Hulse-Taylor binary pulsar system is provided.
In Sec.~\ref{sec:strong}, we study how the theory behaves in the strong gravity regime.

Throughout the paper, we consider a flat Minkowskian metric $\eta_{ab}$ with signature $(-,+,+,+)$. In global inertial coordinates $(ct,x,y,z)$, the line element is
\begin{equation}
ds^2 = -c^2dt^2 +dx^2+dy^2+dz^2\,.
\end{equation}

\section{Self-consistent Lagrangian density} 
\label{sec:lag}

\subsection{A general Lagrangian density}

We consider a general Lagrangian density that couples the gravitational scalar potential $\Phi$ to the energy-momentum tensor, given by
\begin{multline}
\label{eq:genlag}
\mathcal{L}= \mathcal{L}_{\rm matter} -\frac{1}{8 \pi G} f_1\left(\frac{\Phi}{c^2}\right) (\partial_a \Phi)(\partial^a \Phi) +\\
+\beta f_2\left(\frac{\Phi}{c^2}\right) T^{ab} \frac{(\partial_a \Phi)(\partial_b \Phi)}{(\partial_c \Phi)(\partial^c \Phi)}  + f_3\left(\frac{\Phi}{c^2}\right) T \, ,
\end{multline}
where Latin indices run from 0 to 3, $f_1$, $f_2$ and $f_3$ are dimensionless scalar functions satisfying $ f_1(0)=1 $, $f_i(0)=0$ and $f_i'(0)=1$ ($i=2,3$), $T^{ab}$ is the matter and radiation energy-momentum tensor, $T=T^a_a$ is its trace, $\beta$ is a dimensionless constant, $G$ is the Newtonian gravitational constant, and $\partial^a \equiv \eta^{ab} \partial_b$.
Note that the energy momentum tensor couples to the gravitational potential $\Phi$ in two terms, and that $\mathcal{L}_{\rm matter}$ does not depend on $\Phi$. 
To alleviate the notation, we will in this work drop the functions $ f_i $'s arguments.

We keep $c$ in order to track the post-Newtonian order.
In the weak-field, quasi-static limit and taking $c \to \infty$, this Lagrangian reduces to the Newtonian gravity Lagrangian density
\begin{equation}
{\cal L} = - \frac{1}{8 \pi G}|\nabla \Phi|^2 - \rho_m \Phi + \mathcal{L}_{\rm matter} \,,
\end{equation}
where $\rho_m = T^{00}/ c^2$ is the gravitational mass density.

In all generality, we could have let $f_2(0) >0$ in Eq.~\eqref{eq:genlag}, since it would also reduce to Newtonian gravity in the correct weak field limit.
However, as shown in Appendix~\ref{app:extraterm} this would violate special relativity.

\subsection{Accounting for self-interaction: finding $f_1$}

The function $f_1$ describes the field self-interaction. No self-interaction means that $f_1=1$.
We are looking for a Lagrangian density which will make the gravitational coupling to the energy-momentum tensor (EMT) independent of its nature, i.e. it should couple to the gravitational EMT in exactly the same way as any other EMT.

Let us start by calculating the canonical gravitational EMT using Eq.~\eqref{eq:genlag} in a vacuum, that is when $T^{ab}=0$ and $\mathcal{L}_{\rm matter}=0$:
\begin{equation}
\label{eq:vaclag}
\mathcal{L}_{\rm vac} = -\frac{1}{8 \pi G} f_1 (\partial_a \Phi)(\partial^a \Phi) \,.
\end{equation}
The canonical Noether energy-momentum tensor then gives
\begin{align}
T^{ab}_{\rm grav} &\equiv \frac{\partial \mathcal{L}}{\partial(\partial_a \Phi)} (\partial^b \Phi) - \eta^{ab} \mathcal{L} \nonumber \\
\label{eq:gravemt} 
&= - \frac{1}{4 \pi G} f_1  \left((\partial^a \Phi)( \partial^b \Phi) - \frac12 \eta^{ab} (\partial_c \Phi)(\partial^c \Phi) \right) \, ,
\end{align}
and the trace follows immediately,
\begin{equation}
\label{eq:tracemt}
T_{\rm grav} =  f_1 \frac{(\partial_c \Phi)(\partial^c \Phi)}{4 \pi G} \,.
\end{equation}
Note that, since gravity is an attractive force (at least in the Newtonian limit), $T^{00}$ must be negative for a static field (because it requires positive work to pull apart a gravitationally bound system).
Indeed, in such a field, $T^{00}$ is given by 
\begin{equation}
		T^{00}_{\rm grav} =   -\frac{1}{8 \pi G} f_1 \left(\nabla \Phi \right)^2 \,,
\end{equation}
where we used $\eta^{00} = -1$.
Since $f_1 \approx 1$ in the weak field limit, we observe that indeed $T^{00}_{\rm grav}<0$ in this type of configuration. 
This also confirms that $f_1$ must be positive in this limit.

Let us now calculate the full field equations using the Euler-Lagrange formalism.
For a general $T^{ab}$ the field equations are obtained from 
\begin{equation}
	\label{eq:ELeqs}
	\partial_a \left( \frac{\partial \mathcal{L}}{\partial(\partial_a \Phi)}\right) -  \frac{\partial \mathcal{L}}{\partial \Phi} = 0\,,
\end{equation}
and given by
\begin{equation}
	\label{eq:fieldeq}
	\begin{split}
		\square \Phi +  \frac{f_1'}{2 f_1} \frac{w}{c^2} =& -\frac{4 \pi G}{c^2} \left(\frac{f_3' T}{f_1}  + \frac{ \beta f_2'  T^{ab}}{f_1} \frac{(\partial_a \Phi) (\partial_b \Phi)}{w}\right.\\
		&\qquad  \left. -2 \beta  \partial_a \left(\frac{f_2}{f_1} [T\partial\Phi]^a\right)\right)\,,
	\end{split}
\end{equation}
where we defined 
\begin{equation}
\label{eq:divterm}
	[T\partial\Phi]^a \equiv  \frac{T^{ab} \partial_b \Phi - T^{bc}\frac{(\partial_b \Phi) (\partial_c \Phi)}{w}\partial^a \Phi}{w}\,,
\end{equation}
$\square \equiv \partial^a  \partial_a $ is the d'Alambertian, a prime $ ' $ represents the derivative, and $ w \equiv  (\partial^a \Phi) (\partial_a \Phi)$.

To calculate how the gravitational field energy-momentum tensor would contribute in the field equations, we use Eqs.~\eqref{eq:gravemt} and \eqref{eq:tracemt} to plug $T^{ab}_{\rm grav}$ and $T_{\rm grav}$ in the r.h.s. of Eq.~\eqref{eq:fieldeq}. 
Noticing that $[T\partial\Phi]^a=0$ for $T^{ab}_{\rm grav}$, we obtain
\begin{equation}
\label{eq:Tgravcontrib}
\begin{split}
	-\frac{4\pi G}{c^2}\left(f_3'\frac{T_{\rm grav}}{f_1} - f_2' \right.&\left.\frac{\beta T^{ab}_{\rm grav}}{f_1} \frac{(\partial_a \Phi) (\partial_b \Phi)}{w}\right) \\
	&= -\left(f_3' - f_2'\frac{\beta}{2}\right) \frac{w}{c^2}\,.
\end{split}
\end{equation}
In order for the gravitational energy-momentum tensor to source the gravitational potential in the same way as any other type of energy-momentum tensor, the vacuum field equations should have the form
\begin{equation}
\square \Phi = 	-\frac{4\pi G}{c^2}\left(f_3'\frac{T_{\rm grav}}{f_1} - f_2' \frac{\beta T^{ab}_{\rm grav}}{f_1} \frac{(\partial_a \Phi) (\partial_b \Phi)}{w}\right)\,.
\end{equation}
Given that from Eq.~\eqref{eq:fieldeq} the vacuum field equations are given by
\begin{equation}
\square \Phi = -\frac{f_1'}{2 f_1} \frac{w}{c^2}\,,
\end{equation}
we impose that the quantity in Eq.~\eqref{eq:Tgravcontrib} be equal to $-f_1' w/2 f_1 c^2 $. 
We then obtain the following equation relating $f_1$, $ f_2 $, and $ f_3 $:
\begin{equation}
\label{eq:f1expr}
\frac{f_1'}{f_1} = \left(2f_3' - f_2'\beta\right) \Rightarrow f_1 = \exp(2f_3 - \beta f_2)\,.
\end{equation}

Armed with $ f_1 $, which describes the field self-interaction, we are now able to solve the field equations in a vacuum in the weak field limit where both $ f_2 $ and $ f_3 $ are simply $ \Phi/c^2 $.

\section{Field equation in a weak field vacuum}
\label{sec:weak}

Using Eqs.~\eqref{eq:fieldeq} and \eqref{eq:f1expr}, the field equations in a vacuum read
\begin{equation}
\label{eq:fieldeqvacuum}
\begin{split}
	&\square \Phi +  \frac{f_1'}{2 f_1}\frac{w}{c^2} = 0 \\
\Rightarrow\;	&\square \Phi + \frac{2 f_3' - \beta f_2'}{2} \frac{w}{c^2} = 0\,.
\end{split}
\end{equation}
In a static, weak field configuration (first post-Newtonian order), $ \square \Phi \to \nabla^2 \Phi $, $ w \to |\nabla \Phi|^2 $, $ f_3' \to 1 $, and $ f_2' \to 1 $.
Therefore, one gets in this limit
\begin{equation}\label{eq:staticfieldeq}
	\nabla^2 \Phi = -\frac{2- \beta}{2 c^2} |\nabla \Phi|^2 \,.
\end{equation}
Solving Eq.~\eqref{eq:staticfieldeq} in a spherically symmetric configuration around a sphere with mass $M$, we obtain
\begin{equation}
	\label{eq:ssspot}
	\Phi (r) = - \frac{GM}{r} - \frac{2-\beta}{4} \frac{(GM)^2}{c^2 r^2} + \mathcal{O}(c^{-4}) \, ,
\end{equation}
where we imposed the boundary condition $\lim_{r \to \infty} \Phi(r) = 0$, and $GM$ appears in order to recover the Newtonian limit.

\section{Particle Lagrangian and motion}
\label{sec:partlag}

In order to calculate the trajectories predicted by this theory, it is convenient to find the point particle Lagrangian, which is the sum of the free particle and interaction Lagrangians. The special relativistic free particle action $S_{\rm free}$ is just given by 
\begin{equation}
\label{eq:freeaction}
S_{\rm free} = -m_p c^2 \int d\tau =  -m_p c^2 \int dt \frac{d\tau}{dt} \,,
\end{equation}
where $m_p$ is the mass of the particle and $d\tau^2 = -ds^2 = dt^2 - dx^2 -dy^2 -dz^2$. Therefore the free particle Lagrangian is $L_{\rm free} = -m_p c^2 \frac{d\tau}{dt} = -m_p c^2 \sqrt{1-\frac{v^2}{c^2}}$, where $v^2=\boldsymbol{v} \cdot \boldsymbol{v}$, and $\boldsymbol{v}$ is the coordinate 3-velocity. This is just the special relativistic free Lagrangian, as expected. To calculate the interaction Lagrangian, we replace the point particle energy-momentum tensor in the third and fourth terms of the r.h.s. of Eq.~\eqref{eq:genlag}. 
The point particle energy-momentum tensor is given by \cite{WeinbergGravitation1972}
\begin{align}
	\label{eq:pointEMT}
\begin{split}
T^{ab}(\boldsymbol{x},t) &= \rho_0(\boldsymbol{x},t) \, u^a u^b \\
&= m_p \sqrt{1-\frac{v^2}{c^2}} \delta^3(\boldsymbol{x}-\boldsymbol{x}_p(t)) u^a u^b \, ,
\end{split}
\end{align}
where $\rho_0(\boldsymbol{x},t)$ is the proper density, $u^a$ is the 4-velocity, $\boldsymbol{x}_p$ is the position of the particle, and $\boldsymbol{x}_p$ and $\boldsymbol{x}$ are evaluated at the same time $t$. 
Therefore, the particle Lagrangian becomes
\begin{equation}
\label{eq:partlag}
\begin{split}
L_{\rm part} &= L_{\rm free}  +\int d^3 \boldsymbol{x} \left(\beta f_2 T^{ab} \frac{(\partial_a \Phi)(\partial_b \Phi)}{w}  +  f_3 T\right) \\
&=-m_p c^2 \sqrt{1-\frac{v^2}{c^2}}\left(1+ f_3 - \beta f_2 \frac{(u^a \partial_a \Phi)^2}{w\, c^2}\right) \,.
\end{split}
\end{equation}
Using this Lagrangian together with Eq.~\eqref{eq:ssspot} one is able to calculate the trajectories of the planets in the solar system. 

We can also use the particle Lagrangian to derive the gravitational force in the first post-Newtonian order, up to $\mathcal{O}(c^{-4})$. 
In this limit, we can set $f_2 \to \Phi/c^2$ and $f_3 \to \Phi/c^2 + \alpha \Phi^2/c^4$, and the particle Lagrangian becomes
\begin{equation}
\begin{split}
L_{\rm part} &= -m_p c^2 \sqrt{1-\frac{v^2}{c^2}} \\
& \times  \left(1+ \frac{\Phi}{c^2} + \alpha \frac{\Phi^2}{c^4} - \beta \frac{\Phi (\boldsymbol{v} \cdot \nabla \Phi)^2}{(1-v^2/c^2)(\nabla \Phi)^2 c^4}\right)\\
&= -m_p c^2 \sqrt{1-\frac{v^2}{c^2}} \\
& \times \left(1+ \frac{\Phi}{c^2} + \alpha \frac{\Phi^2}{c^4} - \beta \frac{\Phi v^2 \cos^2 \theta }{(1-v^2/c^2)c^4}\right)\,,
\end{split}
\end{equation}
where $\theta$ is the angle between $\boldsymbol{v}$ and $\nabla \Phi$. 

The corresponding Euler-Lagrange equations are:
\begin{widetext}
\begin{equation}
\label{eq:partEOM}
\begin{split}
&\frac{d}{dt}\left(\frac{m_p v^i}{\sqrt{1-\frac{v^2}{c^2}}}\left(1+\frac{\Phi}{c^2} + \alpha \frac{\Phi^2}{c^4} +\beta \frac{\Phi v^2 \cos^2 \theta}{(1-v^2/c^2)c^4} \right) + \frac{2 m_p \beta \Phi}{c^2\sqrt{1-\frac{v^2}{c^2}}} \partial_i \Phi\frac{(\boldsymbol{v}\cdot\nabla\Phi)}{(\nabla \Phi)^2}\right) \\
& \qquad \qquad + m_p \sqrt{1-\frac{v^2}{c^2}} \partial_i\left(\Phi + \alpha \frac{\Phi^2}{c^2}\right) -\beta\frac{m_p v^2}{\sqrt{1-\frac{v^2}{c^2}}c^2}   \partial_i(\Phi \cos^2 \theta) = 0\,,
\end{split}
\end{equation}
\end{widetext}
where $i \in \{x,y,z\}$.
This can be written in a much simpler way using the effective metric formalism as shown in App.~\ref{app:effectivemetric}.

In the second line of Eq.~\eqref{eq:partEOM}, we can see that there are two components of the gravitational force acting on a particle. 
If $v \to c$, only the second component, proportional to $\beta$, acts. 
In the Newtonian limit $c \to \infty$, we recover Newtonian gravitation $\dot{v}^i = -\partial_i \Phi$. 
The canonical momentum (inside the time derivative) has components both proportional to the velocity and to the gravitational field $\partial_i \Phi$. 
Using Eq.~\eqref{eq:partEOM} we can calculate the trajectory of massive particles predicted by the theory, and then compare its predictions to the post-Newtonian tests of general relativity (GR).

Note that in this post-Newtonian limit we have two free parameters: $ \alpha $ and $ \beta $. 
In the next section, we analyze how the weak equivalence principle applied to extended bodies predicts first a relation between these two scalars, and then provides a motivation for their numerical values. 
In this way, we can really have a prediction of the post-Newtonian effects and not only adjust to the observed phenomena.

\section{Constraints from the Equivalence Principle}
\label{sec:ep}

We can use the Equivalence Principle (EP) to put constraints in this theory, as Nordstr\"{o}m did in \cite{Nordstrom1913}. 
In our case, we consider an EP that postulates that the inertial and gravitational masses are equal, independently of composition \cite{Zych2019}.
Throughout this section, for simplicity, we will set $c=1$.

\subsection{Finding a relation between $ \beta $ and $ f_3 $}

The EP states in particular that the gravitational mass $M_G$ is equal to the inertial mass $M_I$ for an extended body. 
The gravitational mass is such that the potential can be written as
\begin{equation}
	\label{eq:phiinertial}
	\tilde{\Phi}(r) = - \frac{G M_G}{r}\,,
\end{equation}
very far away from the source, for suitable definitions of $\tilde{\Phi}$.
We will also define the inertial mass in volume $V$ as \cite{Norton1992}
\begin{equation}
	M_I \equiv \int_V -(T + T_{\rm grav} + T_{\rm int}) \,dV\,,
\end{equation}
where $ T $, $ T_{\rm grav} $, and $ T_{\rm int} $ are traces of the matter, gravity, and interaction energy-momentum tensors. $ T^{ab}_{\rm int} $ is obtained from th Lagrangian density in Eq.~\eqref{eq:genlag} and given in general by
\begin{multline}
	T^{ab}_{\rm int} = -\eta^{ab}  \left(f_3 T + \beta f_2 T^{cd}\frac{(\partial_c \Phi)(\partial_d \Phi)}{w}\right)\\
	+ 2\beta f_2 [T \partial \Phi]^a \partial^b \Phi\,,
\end{multline}
where the vector $ [T \partial \Phi]^a $ is defined in Eq.~\eqref{eq:divterm}.

To obtain an equation like Eq.~\eqref{eq:phiinertial}, one can define $\tilde{\Phi} = c^2 g(\Phi/c^2)$, with $ g' = \sqrt{f_1} $ and $ g(0) = 0 $ such that
\begin{equation}
	\label{eq:phitilde}
	\square \Phi + \frac{f_1'}{2 f_1} \frac{w}{c^2} = \frac{\square \tilde{\Phi}}{\sqrt{f_1}}\,,
\end{equation}
and 
\begin{equation}
	w = \frac{\tilde{w}}{f_1}\,,
\end{equation}
where $ \tilde{w} \equiv (\partial_a \tilde{\Phi})  (\partial^a \tilde{\Phi})$.
From now on in this section we will consider a static configuration  (i.e. only $T^{00} = \rho c^2 \neq 0$ and $\partial_0 \Phi = 0$).
In such a configuration, one can solve Eq.~\eqref{eq:fieldeq} with $ \tilde{\Phi} $, obtaining 
\begin{equation}
	\nabla^2 \tilde{\Phi} = -f_3'\frac{4 \pi G T}{c^2\sqrt{f_1}} =  f_3'\frac{4 \pi G \rho}{\sqrt{f_1}}\,.
\end{equation}
Very far away from the source $T$, one obtains from Gauss' Law
\begin{equation}
	|\nabla \tilde{\Phi}| = \frac{G}{r^2}  \int_V  f_3'\frac{\rho}{\sqrt{f_1}}\,,
\end{equation}
from where we read the gravitational mass:
\begin{equation}
	\label{eq:mg}
	M_G = \int_V  f_3'\frac{\rho}{\sqrt{f_1}}\,.
\end{equation}

Let us now calculate the inertial mass. We are using $T^{ab} = \rho c^2 \delta_0^a  \delta_0^b$, implying $ T = -\rho c^2$. $T_{\rm grav}$ is given in Eq.~\eqref{eq:tracemt}. In our static configuration, we can also calculate the $ T^{ab}_{\rm int} $, obtaining $ T^{ab}_{\rm int} = - \eta^{ab} T f_3 $. Taking the trace, we obtain
\begin{equation}
	T_{\rm int} = 4 \rho f_3\,.
\end{equation}
So, putting all together, we have
\begin{equation}
	M_I = \int_V \left( -\frac{\tilde{w}}{4 \pi G c^2} + \rho - 4 \rho f_3 \right)\,.
\end{equation}
Let us focus in the simplifying the first term. 
For that, consider the following equality
\begin{equation}
	\begin{split}
		\int_V |\nabla \tilde{\Phi}|^2 &dV \\
		&= \int_S \tilde{\Phi} (\nabla \tilde{\Phi} \cdot \vec{n}) \,dS - \int_V \tilde{\Phi} \nabla^2 \tilde{\Phi}\, dV\,,
	\end{split}
\end{equation}
where $S$ is the boundary of volume $V$ and $\vec{n}$ is the outward pointing unit normal.
If $\tilde{\Phi} = \mathcal{O}(1/r) $, $ \int_S \tilde{\Phi} (\nabla \tilde{\Phi} \cdot \vec{n}) \,dS $ goes to 0 as we increase $V$.
Taking a spherical region, much larger than the source, we then have
\begin{equation}
	\int_V |\nabla \tilde{\Phi}|^2 dV = - \int_V \tilde{\Phi} f_3'\frac{4 \pi G \rho}{c^2\sqrt{f_1}}\, dV\,,
\end{equation}
and so we obtain
\begin{equation}
	\label{eq:mi}
	\begin{split}
		M_I &= \int_V \left( \tilde{\Phi} f_3'\frac{\rho}{c^2\sqrt{f_1}} + \rho - 4 \rho f_3 \right)\\
		&= \int_V \rho \left( \frac{\tilde{\Phi} f_3'}{c^2\sqrt{f_1}} + 1 - 4 f_3 \right)\,.
	\end{split}
\end{equation}
We can now use Eqs.~\eqref{eq:mg} and \eqref{eq:mi} and set $M_G = M_I $, getting 
\begin{equation}
	\int_V  f_3'\frac{\rho}{\sqrt{f_1}} = \int_V \rho \left( \frac{\tilde{\Phi} f_3'}{c^2\sqrt{f_1}} + 1 - 4 f_3 \right)\,.
\end{equation}
To guarantee that the equality holds for any mass distribution $\rho$, we need 
\begin{equation}
	\frac{f_3'}{\sqrt{f_1}} = \frac{\tilde{\Phi} f_3'}{c^2\sqrt{f_1}} + 1 - 4 f_3\,.
\end{equation}
To simplify this condition, notice that $f_3'/\sqrt{f_1} = c^2 df_3/d\tilde{\Phi} $. So we get the following differential equation for $ f_3 $
\begin{equation}
	c^2\frac{df_3}{d\tilde{\Phi}} (1 -  \frac{\tilde{\Phi}}{c^2}) + 4 f_3 =  1 \,,
\end{equation}
which can be solved to get
\begin{equation}
	\label{eq:f3tilde}
	f_3 = \frac{1}{4} + c_1 \left(1 - \frac{\tilde{\Phi}}{c^2}\right)^4\,,
\end{equation}
where $c_1$ is an integration constant. 
Given that $ f_3 (\Phi = 0) = 0 $ and $ \tilde{\Phi} = \Phi +\mathcal{O}(\Phi^2) $, we get $ c_1 = -1/4 $. 
To see what conditions the EP assumption forces in the function $ f_3 $, we have to write it as a function of $ \Phi $ and not $ \tilde{\Phi} $.
For that, recall that $ d\tilde{\Phi}/d\Phi = \sqrt{f_1} $, and using Eq.~\eqref{eq:f1expr} we have
\begin{equation}\label{eq:phitildeapprox}
	\frac{d\tilde{\Phi}}{d\Phi} = \exp\left(f_3 - \frac{\beta}{2} f_2\right)\,,
\end{equation}
and given that
\begin{align}
	f_2 &= \frac{\Phi}{c^2} + \mathcal{O}\left(\frac{\Phi^2}{c^4}\right)\,,\\
	f_3 &= \frac{\Phi}{c^2} + \mathcal{O}\left(\frac{\Phi^2}{c^4}\right)\,,
\end{align}
we have 
\begin{equation}
	\exp\left(f_3 - \frac{\beta}{2} f_2\right) = 1 + \left(1-\frac{\beta}{2}\right)\frac{\Phi}{c^2} + \mathcal{O}\left(\frac{\Phi^2}{c^4}\right)\,.
\end{equation}
Note that we don't have to go to second order in $ \Phi $ because this is $ d\tilde{\Phi}/d\Phi $ and not $ \tilde{\Phi} $ directly.
We can then integrate, getting 
\begin{equation}
	\tilde{\Phi} = \Phi + \left(\frac{1}{2}-\frac{\beta}{4}\right) \frac{\Phi^2}{c^2} + \mathcal{O}\left(\frac{\Phi^3}{c^6}\right)\,,
\end{equation}
and we can use this expression to expand $ f_3 $ in Eq.~\eqref{eq:f3tilde} to second order in $ \Phi $, obtaining:
\begin{equation}\label{eq:f3final}
	f_3(\Phi) = \frac{\Phi}{c^2} - \left(1 + \frac{\beta}{4}\right) \frac{\Phi^2}{c^4} + \mathcal{O}\left(\frac{\Phi^3}{c^6}\right)\,.
\end{equation}
So interestingly the EP implies a relation between $ f_3 $ and $ \beta $.

\subsection{Fixing $ \beta $ with the EP}

We can put a constraint in $ \beta $ using the EP and comparing the accelerations  in a weak field for massless and massive particles. 
Let us start with massless particles.
We work in the massless limit of Eq.~\eqref{eq:partEOM} where $m_p \to 0 $, $ v \to c $, and $ \frac{m_p}{\sqrt{1-\frac{v^2}{c^2}}} \to E/c^2 \equiv m_\gamma$. Furthermore, we are also in the very weak field limit, specifically given $v^2 = c^2(1-\epsilon)$, then we have $ \Phi/c^2 = \mathcal{O}(\epsilon^2)$.

Sending $ \epsilon \to 0 $, the equation of motion then becomes
\begin{equation}
	\frac{d}{dt}\left( m_\gamma v^i \right) 
	-\beta m_\gamma \partial_i(\Phi \cos^2 \theta) = 0\,,
\end{equation}
where $ \theta $ is the angle between $ v^i $ and $ \partial_i \Phi $. 

For slowly moving massive particles, we have the following simple equations of motion
\begin{equation}
	\frac{d}{dt}\left( m_p v^i \right) + m_p \partial_i\Phi  = 0\,.
\end{equation}

Now, suppose we had a massless box of photons, and wanted to calculate the total gravitational pull on the photons inside that box. 
Assuming the gravitational field is weak and constant in the box, each photon feels an acceleration given by $ \beta \partial_i(\Phi \cos^2 \theta) $. 
If there is no preferred direction in the velocities of the photons, and if the velocity for a single photon does not vary much, on average the acceleration will be $ \frac{1}{2}\beta \partial_i\Phi $. 
To satisfy the EP, i.e. if we set the acceleration of photon box equal to the acceleration of the massive slowly moving particle, we must have
\begin{equation}\label{eq:wep}
	\frac{1}{2}\beta \partial_i\Phi = -\partial_i\Phi \Rightarrow \beta = -2\,.
\end{equation} 
Now, admittedly, if we were considering a beam of photons in a particular direction, the factor $ \cos^2 \theta $ would cause the two accelerations to be different.
However, this condition (taking the average of $ \cos^2 $) is the closest we can get to the EP.
Indeed, it is hard to understand how one can obtain an Equivalence Principle like the one general relativity from a theory on a Minkowski spacetime.
Note that this direction dependent gravitational acceleration for weak fields also happens in general relativity \cite{McGruder1982}.

\subsection{Combining the two conditions}

These two conditions on $ f_3 $ and $ \beta $ can be combined as
\begin{align}
	\label{eq:betawep}
	&\beta = -2 \,,\\
	&f_3 = \Phi - \left(1+\frac{\beta}{4}\right) \frac{\Phi^2}{c^2} + \mathcal{O}(\Phi^3)\,.
\end{align}
Defining $ \alpha = - \left(1+\frac{\beta}{4}\right) $, we have $ f_3 = \Phi + \alpha \Phi^2 + \mathcal{O}(\Phi^3) $. 
Fixing $ \beta $ as in Eq.~\eqref{eq:betawep}, we get 
\begin{equation}
	\label{eq:alphawep}
	\alpha = -\frac{1}{2}\,.
\end{equation}

In this section, we provided a motivation for the values of parameters $ \alpha $ and $ \beta $ using the WEP.
Admittedly, the relation given in Eq.~\eqref{eq:f3final} is more robust than the numerical values themselves, but these estimates will allow us to give definite predictions for the theory. 

In the next section, we use the PPN formalism to show that this theory can predict the outcome of all classical tests of GR.

\section{PPN limit of the theory} \label{sec:ppn}

To compare this theory with general relativity, it is convenient to find its PPN parameters. 
However, as this is a field theory in Minkowski spacetime, and not a metric theory of gravity, it does not fit into the standard PPN paradigm. 
Therefore, we have to adapt and find an effective metric corresponding to the particle Lagrangian given in Eq.~\eqref{eq:partlag}, in the post-Newtonian limit. 

\subsection{Effective metric}

Looking back to Eq.~\eqref{eq:partlag}, one can extract an effective metric that explains the trajectories of particles.
This approach will be similar to the one in \cite{Novello2012} in that the trajectories of the particles can be seen as geodesics of a metric different from the background.
However here the effective metric comes as an emergent phenomenon from the Lagrangian density as explained in \cite{Braganca2018} and is not given \textit{a priori}.
Let us see how one can do that.
First, we define a new function $f$ as 
\begin{equation}
f \equiv 1+ f_3 - \beta f_2 \frac{(u^a \partial_a \Phi)^2}{w\, c^2} \,,
\end{equation}
such that the Lagrangian in Eq.~\eqref{eq:partlag} becomes 
\begin{equation}
\label{eq:partlag2}
L_{\rm part} = -m_p c^2 \sqrt{1-\frac{v^2}{c^2}} f \,.
\end{equation}
Following \cite{Braganca2018}, the effective metric corresponding to the point particle Lagrangian in Eq.~\eqref{eq:partlag2} (that is, the metric whose geodesics are the trajectories described by the Lagrangian \eqref{eq:partlag2})  is simply
\begin{align}
	\label{eq:effmetric}
	ds_{\rm eff}^2 &= -f^2 d\tau^2 =  f^2 \, \eta_{ab} \, dx^a \, dx^b \,.
\end{align}
Note that in general $f$ depends on the 4-velocity of the particle $u^a$, which just means that this effective metric will too depend on the local velocity of the particle.
In Appendix~\ref{app:effectivemetric}, we show that to first PPN order and for a static field, $ds_{\rm eff}^2$ can be obtained from an effective metric $g_{ab}$ given in general by
\begin{multline}
	g_{ab} = \left(1+ \frac{2\Phi}{c^2} + (2\alpha+1) \frac{\Phi^2}{c^4}\right)\eta_{ab}\\
	+ 2\beta \frac{\Phi}{c^2} \frac{(\partial_a \Phi)(\partial_b \Phi)}{w}\,,
\end{multline} 
such that 
\begin{equation}
\label{eq:dseffgab}
	ds_{\rm eff}^2 = g_{ab} dx^a dx^b\,.
\end{equation}
To get this result, we used that $ f_2 = \Phi/c^2 $, $f_3 = \Phi/c^2 + \alpha \, \Phi^2/c^4$ to this order.
In particular, to this order the effective metric is independent of the 4-velocity, which makes the comparison with the standard PPN formalism very convenient (this is no longer true at higher orders and so the analysis becomes more subtle). 
According to Eq.~\eqref{eq:dseffgab}, all lengths and times measured using physical objects (that follow geodesics too) will be measured in terms of this effective metric, possibly corresponding to a curved spacetime, and not in terms of the flat Minkowski metric.
So, from a purely scalar theory of gravity on a flat spacetime, we obtain that the trajectories of point particles are geodesics of an effective metric that is emergent from the Lagrangian density and not postulated \textit{a priori}.

We want to calculate the first post-Newtonian order of this theory, and its applications to the solar system tests, where the gravitational field generated by the Sun can be considered weak, static and spherically symmetric thus given by Eq.~\eqref{eq:ssspot}. 
Therefore, we get for $ds_{\rm eff}^2$ in this regime using spherical coordinates
\begin{multline}
	ds_{\rm eff}^2 = -c^2 \left[ 1- \frac{2m}{r} + \left(\frac{\beta}{2} + 2\alpha \right) \frac{m^2}{r^2} \right] dt^2  \\
	+ \left( 1- \frac{2m}{r} (1 + \beta) \right) dr^2 + \left( 1- \frac{2m}{r} \right) r^2 d\Omega^2 \,,
\end{multline}
This metric cannot be used to read the PPN parameters yet. 
For that, one needs to write it in isotropic coordinates.

Introducing another radial variable $\rho$ defined as
\begin{equation}
\label{eq:rho}
\rho \equiv r + \beta m \,,
\end{equation}
we get, to first post-Newtonian order \cite{Will1972},
\begin{multline}
\label{eq:finalmetric}
ds_{\rm eff}^2 = -c^2 \left[ 1 - \frac{2m}{\rho} +  \left(- \frac{3 \beta}{2} + 2\alpha \right) \frac{m^2}{\rho^2} \right] dt^2 \\
+ \left( 1- \frac{2m}{\rho}(1+\beta)\right) \left(d\rho^2 + \rho^2 d\Omega^2 \right) \,,
\end{multline}
which is exactly what we need to read the PPN parameters.

\subsection{PPN parameters}

For a static, spherically symmetric spacetime, the PPN metric \cite{Will1972} is given by 
\begin{equation}
\label{eq:PPN}
\begin{split}
ds^2 = - &\left( c^2 - 2 \frac{GM}{r} + 2 \beta_{\rm PPN} \frac{(GM)^2}{c^2 r^2} \right) dt^2 +  \\
 &+ \left(1 + 2 \gamma_{\rm PPN} \frac{GM}{c^2 r} \right) (dr^2 + r^2 \, d\Omega^2) \,,
 \end{split}
\end{equation}
where $\beta_{\rm PPN}$ and $\gamma_{\rm PPN}$ are the relevant PPN parameters in this case. 
For general relativity, $\beta_{\rm PPN}=\gamma_{\rm PPN}=1$, and all solar system test predictions (including the so-called classical tests of GR) depend on $\beta_{\rm PPN}$ and $\gamma_{\rm PPN}$.

Using Eq.~\eqref{eq:finalmetric}, we get the following PPN parameters $\beta_{\rm PPN}$ and $\gamma_{\rm PPN}$,
\begin{align}
\label{eq:betappn}
\beta_{\rm PPN} = -\frac{3 \beta}{4} + \alpha \\
\label{eq:gammappn}
\gamma_{\rm PPN} = - ( 1 + \beta) \,.
\end{align}
In order to recover the correct results for the classical tests, we just have to get $\beta_{\rm PPN}=\gamma_{\rm PPN}=1$.
It is remarkable that the predictions for $ \beta $ and $ \alpha $ using the EP in Eqs.~\eqref{eq:wep} and \eqref{eq:alphawep} are exactly the values needed to obtain the PPN parameters corresponding to GR, that is
\begin{align}
\label{eq:beta}
\beta &= -2 \,,\\
\alpha &= -\frac{1}{2} \,.
\end{align}
From Eq.~\eqref{eq:gammappn} it is clear that if $\beta=0$ (that is if we removed the term proportional to $T^{ab} \partial_a\Phi \partial_b\Phi$ in Eq.~\eqref{eq:genlag}), it would be impossible to obtain the correct PPN parameters. 
That is why, for example, Nordstr\"{o}m's theory of gravity could not explain the classical solar system tests.

\begin{figure}[h]
	\centering
	\includegraphics[width=\columnwidth]{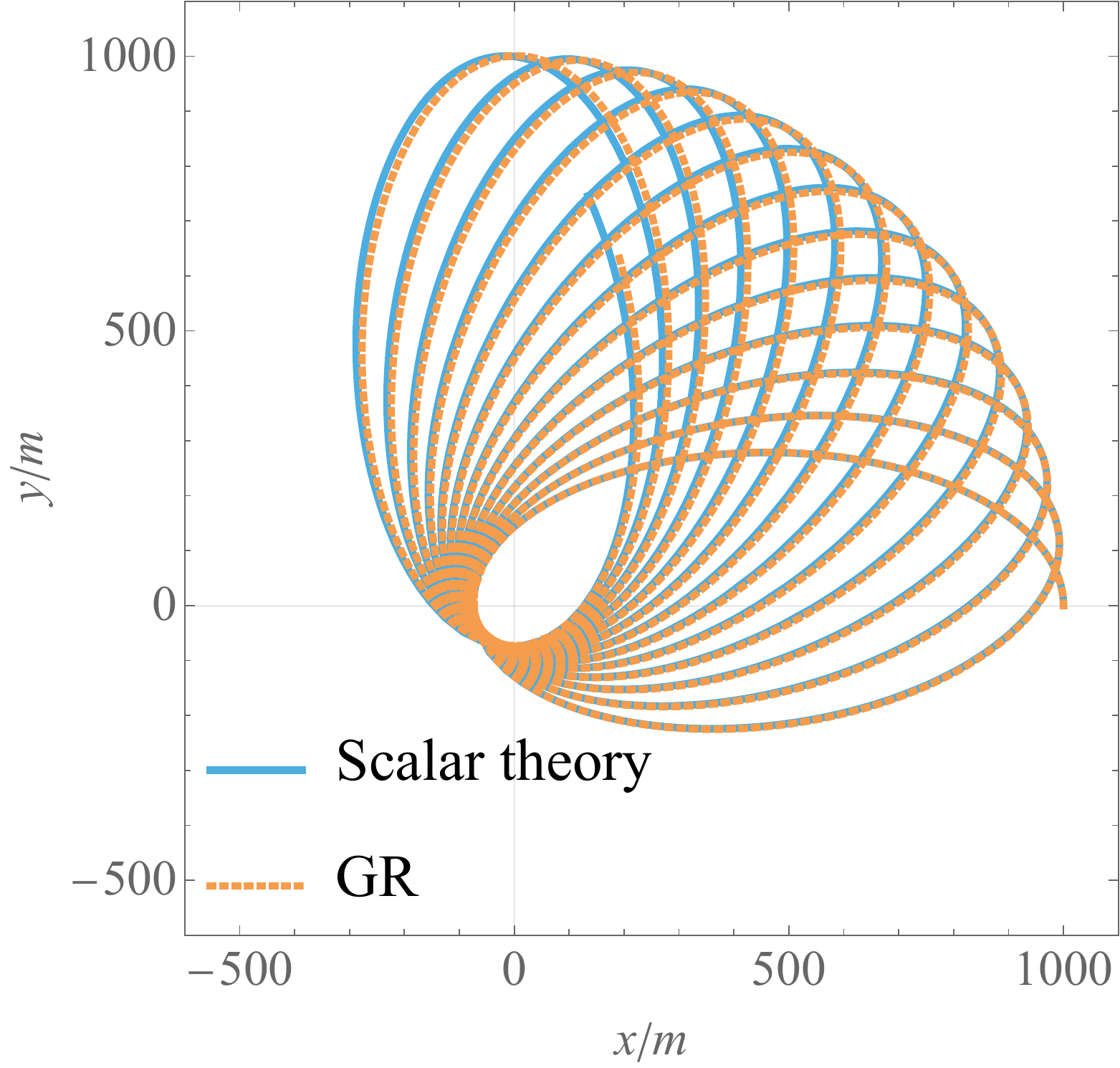}
	\caption{Orbits for a test particle around a mass located at the origin for $ \alpha = -\frac{1}{2} $ and $ \beta = -2 $. 
	The unit in each axis is the geometrical mass of the central mass $m = GM/c^2$.
	The precession is equal at the first post Newtonian level, but there are higher order differences that cause a small lag after several periods.} 
	\label{fig:orbit}
\end{figure}

So far, we have rigorously analyzed the behavior of a massive point particle. 
This theory can then correctly explain Mercury's perihelion precession.
In fact, one can solve the Euler-Lagrange equations using the Lagrangian in Eq.~\eqref{eq:partlag} for the scalar theory and compare the orbit of a planet in general relativity, verifying that they match even in the case of mildly relativistic velocities (1\% of the speed of light). This is shown in Fig.~\ref{fig:orbit}.
Note that this matching with GR is obtained using directly the Lagrangian in Eq.~\eqref{eq:partlag} and not using the effective metric. 
This confirms that the trajectories predicted by this theory can be interpreted as geodesics of a curved spacetime.
However, it is not obvious that these conclusions are also applicable to electromagnetic fields (i.e. photons). 
We could take the limit of this approach when $ v \to c $, but that limit is very tricky to calculate.
To be certain that this theory can explain the bending of light and the Shapiro delay, one needs to describe rigorously the behavior of light. 
That is the next section's goal.

\section{Behavior of light}
\label{sec:light}

We follow a similar approach as in the last section to study the propagation of light. 
The strategy will be first to write the Lagrangian for the electromagnetic (EM) field in the PPN formalism, then to find the corresponding Lagrangian in our scalar theory, and finally to check if they agree for the parameters previously found under the coordinate transformation \eqref{eq:rho}.

In GR, the EM Lagrangian in a vacuum is given by
\begin{align}
\label{eq:GRemLag}
\mathcal{L}^{\rm GR}_{\rm EM} &= -\frac{1}{4 \mu_0} F_{ab} F^{ab} \frac{\sqrt{-g}}{c} \\
&= -\frac{1}{4 \mu_0} F_{ab} F_{cd} \,g^{ac}g^{bd} \frac{\sqrt{-g}}{c} \,,
\end{align}
where $F_{ab} \equiv \partial_a A_b - \partial_b A_a$, $A_a$ is the 4-potential, $g^{ab}$ is the inverse metric, and $g$ is the determinant of the metric.

In the Lagrangian given in Eq.~\eqref{eq:genlag}, we need to replace $T^{ab}$ by the EM energy momentum tensor, given by
\begin{equation}
T^{ab} = \frac{1}{\mu_0} \left(F^{ac} F^b_c - \frac{1}{4} \eta^{ab} F^{cd}F_{cd} \right)\,.
\end{equation}
Since the metric is flat, the indices are raised and lowered using the Minkowski metric $\eta_{ab}$. The trace of the EM energy momentum  tensor vanishes, so that the Lagrangian $ \mathcal{L}^{\Phi}_{\rm EM} $ for light in the scalar theory to first post-Newtonian order becomes
\begin{align}
\mathcal{L}^{\Phi}_{\rm EM} &= \beta T^{ab} \frac{(\partial_a \Phi)(\partial_b \Phi)}{w} \frac{\Phi}{c^2} -\frac{1}{4 \mu_0} F_{ab} F^{ab} \,.
\end{align}
Assuming a radial $\Phi$ given by Eq.~\eqref{eq:ssspot}, and spherical coordinates $x^a = (t,r,\theta,\phi)$, one obtains
\begin{equation}
\begin{split}
\mathcal{L}^{\Phi}_{\rm EM} &= -\frac{1}{4 \mu_0}\sqrt{-\eta} F_{ab} F_{cd} \times\\ 
&\times \left(\eta^{ac}\eta^{bd} (1+\beta\Phi/c^2) - \eta^{ac} h^{bd} - \eta^{bd} h^{ac} \right) \,,
\end{split}
\end{equation}
where $h^{ab}$ is given by
\begin{equation}
\begin{split}
h^{ab} = 2 \beta \frac{\Phi}{c^2} \frac{(\partial^a \Phi)(\partial^b \Phi)}{w}\,.
\end{split}
\end{equation}
Note that if $\beta = 0$, then as expected the Lagrangian becomes the EM Lagrangian in flat Minkowski space without gravity. 
Doing the change of variable defined in Eq.~\eqref{eq:rho} that enables us to compare with the PPN metric, we find that $\mathcal{L}^{\rm GR}_{\rm EM} = \mathcal{L}^{\Phi}_{\rm EM} + \mathcal{O}(m^2)$, provided that $\beta = -2$, which agrees with the value found for $\beta$ in Eq.~\eqref{eq:beta}. 
This confirms that this scalar theory of gravity agrees with the bending of light predicted by GR. Thus, this scalar theory explains all the classical tests of General Relativity, making it a classically viable and consistent field theory of gravity.

Having showed that this theory is valid and agrees with GR in the first post-Newtonian order, it is of course interesting to analyze this relativistic scalar theory of gravity in the strong field regime, and compare its predictions with general relativity.

\section{Gravitational radiation}
\label{sec:gw}

From Eq.~\eqref{eq:phitilde}, it is clear that gravitational waves propagate at the speed of light in a vacuum, confirming the results from gravitational wave astronomy.
We will in this section use $\tilde{\Phi}$ instead of $\Phi$ to simplify calculations.
Let us analyze the generation of gravitational waves, in a system like the Hulse-Taylor binary pulsar~\cite{Weisberg2004}.
We will follow a similar approach to the one used in \cite{Shapiro1993} and \cite{Landau1975}.
Using Eq.~\eqref{eq:fieldeq} with $\tilde{\Phi}$, we obtain:
\begin{equation}
\begin{split}
		\frac{\square \tilde{\Phi}}{\sqrt{f_1}} =& -\frac{4 \pi G}{c^2} \left(\frac{f_3' T}{f_1}  + \frac{ \beta f_2'  T^{ab}}{f_1} \frac{(\partial_a \tilde{\Phi}) (\partial_b \tilde{\Phi})}{\tilde{w}}\right.\\
		&\qquad  \left. -2 \beta  \partial_a \left(\frac{f_2}{f_1} [T\partial\Phi]^a\right)\right)\,,
\end{split}
\end{equation}
and using the Green's function approach (analogously to electromagnetism), we can write the solution as
\begin{equation}
\begin{split}
	\tilde{\Phi}(t,\vec{x}) =& \frac{G}{c^2} \int_V \frac{d^3 \vec{x}'}{|\vec{x}-\vec{x}'|} \times\\
	&\times \left(\frac{f_3' T}{\sqrt{f_1}}  + \frac{ \beta f_2'  T^{ab}}{\sqrt{f_1}} \frac{(\partial_a \tilde{\Phi}) (\partial_b \tilde{\Phi})}{\tilde{w}}\right)_{t_{\rm ret}}\,,
\end{split}
\end{equation}
where $\vec{x}$ is the observer's position, we evaluate the integrand at the retarded time $t_{\rm ret} = t - |\vec{x}-\vec{x}'|/c$, and we neglected the divergence term because it becomes a surface term, where $T^{ab}=0$.
In the wave zone, considering that the coordinate's origin is the center of mass of the system we analyze, we can set the denominator to $|\vec{x}|$.
We can now use the expression of the dust energy momentum tensor in Eq.~\eqref{eq:pointEMT}, obtaining
\begin{equation}
	\label{eq:wavezone}
	\begin{split}
		\tilde{\Phi}(t,\vec{x}) =& -\frac{G}{|\vec{x}|} \int_V d^3 \vec{x}'\frac{\rho_0(\vec{x}')}{\sqrt{f_1}}
		\left(f_3' - \beta f_2' \frac{(u^a \partial_a \tilde{\Phi})^2}{c^2\tilde{w}} \right)_{t_{\rm ret}}\,,
	\end{split}
\end{equation}
where $u^a$ also depends on $\vec{x}'$.
Now, for dust we can write $\rho_0(\vec{x}')$ as $\sqrt{1-v^2/c^2}\rho_m(\vec{x}')$, where $\rho_m(t,\vec{x}) = \sum m_p \delta^3(\vec{x}-\vec{x}_p(t))$ is the mass density for an observer far away. 
So, in the first post-Newtonian order, one has
\begin{equation}
	\rho_0 \approx \rho_m \left(1 - \frac{v^2}{2 c^2}\right)\,.
\end{equation}
Moreover, at this order we have from Eq.~\eqref{eq:f3tilde} $f_3'/\sqrt{f_1} = 1 - 3\tilde{\Phi}/c^2$ and $f_2'=1$.
One can simplify the numerator of the term proportional to $\beta$ in the following way (in the first post-Newtonian order):
\begin{equation}
u^a \partial_a\tilde{\Phi} \approx u^a \dot{u}_a = \frac{d}{d\tau}\left(\frac{u^a u_a}{2}\right) = 0\,,
\end{equation} 
where $\tau$ is the proper time measured with the Minkowski metric along the particle's trajectory, and here dot means differentiation with respect to $\tau$. 
We shall then neglect this term at this order.
So, Eq.~\eqref{eq:wavezone} becomes
\begin{equation}
	\label{eq:wavezone2}
	\begin{split}
		\tilde{\Phi}(t,\vec{x}) =& - \frac{G}{|\vec{x}|} \int_V d^3 \vec{x}'\rho_m\left(1-\frac{v^2}{2c^2} - 3 \frac{\tilde{\Phi}}{c^2}\right)_{t_{\rm ret}}\,.
	\end{split}
\end{equation}
To analyze this expression for a binary system, we can expand $\rho_m$ in the following way:
\begin{equation}
\label{eq:rhomexpand}
\begin{split}
\rho_m(t_{\rm ret}, \vec{x}') = \rho_m(t - \frac{r}{c}, \vec{x}')& + \frac{(r - |\vec{x}-\vec{x}'|)}{c} \rho_{m,t}\\
+& \frac{1}{2} \frac{(r - |\vec{x}-\vec{x}'|)^2}{c^2} \rho_{m,tt}\,,
\end{split}
\end{equation}
where we defined $r\equiv|\vec{x}|$.
The first term represents the non-radiative gravitational potential, so we drop it.
In the wave zone (large $r$), we can simplify 
\begin{equation}
r - |\vec{x}-\vec{x}'| \approx \frac{\vec{x}\cdot\vec{x}'}{r}\,.
\end{equation}
Let us know specify a system of point particles labeled by $p$. 
Then the second term of Eq.~\eqref{eq:rhomexpand}, when inside the integral to first post-Newtonian order, can be written as 
\begin{equation}
\begin{split}
	\frac{\vec{x}}{c r} \int_V d^3 \vec{x}' \partial_t x'_i \rho_m = 	\frac{x^i}{c r} \sum_p m_p \partial_t \vec{x}_p \,,
\end{split}
\end{equation}
but $\sum_p m_p \partial_t \vec{x}'_p$ is just the total momentum of the system, which is constant, and in particular can be taken to be zero if the observer is not moving relative to the center of mass.
Therefore we can neglect this term too.

So we can now focus on the third term in Eq.~\eqref{eq:rhomexpand}. 
To first post-Newtonian order, we can write the contribution of that term in the integral Eq.~\eqref{eq:wavezone2} as
\begin{equation}
	\label{eq:quadrupole}
	\begin{split}
		\tilde{\Phi}(t,\vec{x}) =& -\frac{G}{r} \int_V d^3 \vec{x}'\frac{1}{2} \frac{(\vec{x}\cdot\vec{x}')^2}{c^2 r^2} \rho_{m,tt}\\
		=&  -\frac{G}{2 c^2 r^3} x_\alpha x_\beta\, \partial^2_t \sum_p m_p x^\alpha_p x^\beta_p\,,
	\end{split}
\end{equation}
where Greek letters run from 1 to 3 and represent spatial coordinates, and we recognize the quadrupole moment 
\begin{equation}
	Q^{\alpha\beta} = \sum_p m_p x^\alpha_p x^\beta_p\,.
\end{equation}
So, we can write the radiative $\tilde{\Phi}$ as simply
\begin{equation}
	\tilde{\Phi}(t,\vec{x}) = - \frac{G}{2 c^2 r^3} x_\alpha x_\beta\, \ddot{Q}^{\alpha\beta}\,.
\end{equation}
So we recover a gravitational quadrupole radiation very similar to GR.
We can now calculate the gravitational energy flux using Eq.~\eqref{eq:gravemt}. 
Following \cite{Landau1975}, we can start by calculating the flux in for a wave propagating in the $x$ direction.
To simplify, we will return to $\Phi$, noting that $\partial_a \Phi \partial_b \Phi = \partial_a \tilde{\Phi} \partial_b \tilde{\Phi}/f_1$. 
We then have 
\begin{equation}
\label{eq:T01}
\begin{split}
	c T^{01}_{\rm grav} &= \frac{1}{4 \pi G} f_1  (\partial_t \Phi)( \partial_x \Phi)= -\frac{1}{4 \pi G c} f_1  (\partial_t \Phi)^2\\
	&= -\frac{G}{16\pi  c^5 r^2} \left(n_\alpha n_\beta\, \dddot{Q}^{\alpha\beta}\right)^2\,,
\end{split}
\end{equation}
where in the second equality we used that for a propagating wave $\Phi$ is a function of $t-x/c$, and we defined $n_\alpha \equiv x_\alpha/r$.
Notice that this is a negative quantity, because in this theory the gravitational field has a negative energy density as mentioned in Sec.~\ref{sec:lag}. 
So interestingly gravitational waves carry negative energy, leaving the system with more energy.

Before proceeding, let us comment briefly about this result. 
If we just used the formula for classical orbital energy $E_{\rm orb} = - Gm_1 m_2/2 a$, $a$ being the semi-major axis, then the fact that gravitational waves carry negative energy would cause $a$ to increase, in order for the orbital energy to increase as well. 
This would strongly contradict observations, immediately disproving this theory.
However, to be precise we need to use Noether's theorem for the terms in the Lagrangian Eq.~\eqref{eq:genlag} that depend on $\Phi$ or $\partial \Phi$, which states
\begin{equation}
	\label{eq:emtconservation}
	\partial_a T_{\rm tot}^{ab} = 0\,,
\end{equation}
where $T^{ab}_{\rm tot} = T^{ab}_{\rm int} + T^{ab}_{\rm grav}$.
In particular, using $b=0$ in Eq.~\eqref{eq:emtconservation}, and integrating over a large spherical region around the binary system (so that surface terms vanish), we can approximate
\begin{align}
& \int_V T^{00}_{\rm grav} = -\int_V\frac{1}{8\pi G} |\nabla \Phi|^2 = \frac{1}{2} \int_V \rho_m \Phi \,,\\
&  \int_V T^{00}_{\rm int} = \int_V -\eta^{00} T f_3 = \int_V -\rho_m \Phi \,.
\end{align}
In particular, the sum of the energy stored in the gravitational field and the ``interaction'' energy is non-negative.  
So, given that the masses of each object do not change, Noether's theorem becomes
\begin{equation}
	\frac{d}{dt} \int_V \left(-\frac{\rho_m \Phi}{2}\right) = \int_S n_i T^{0i}_{\rm grav} \,,
\end{equation}
and since the Newtonian gravitational energy is equal to $\int_V \rho_m \Phi/2$, the change of that energy is $\int_S n_i (-T^{0i}_{\rm grav}) = \int_S n_i |T^{0i}_{\rm grav}|>0$.
So, the total Newtonian orbital energy actually decreases, meaning that the two masses in the system spiral toward each other. 
To calculate how much that energy decreases, we have to use $|T^{0i}_{\rm grav}|$ instead of just $T^{0i}_{\rm grav}$. 
Let us now proceed with the calculation from Eq.~\eqref{eq:T01}.

The total (absolute value of the) gravitational energy radiated $dI$ in the solid angle $d\Omega$ is then
\begin{equation}
\label{eq:dI}
\begin{split}
	dI &= c |T^{01}_{\rm grav}| r^2 d\Omega\\
	&= \frac{G}{16\pi  c^5} \left(n_\alpha n_\beta\, \dddot{Q}^{\alpha\beta}\right)^2 d\Omega\,,
\end{split}
\end{equation}
which as expected is independent of $r$.
This result is very similar to the one from general relativity~\cite{Landau1975}. 

Now, as in \cite{Peters1963}, let us consider a system of two particles of masses $m_1$ and $m_2$ in a Keplerian orbit. 
Let us choose as our origin the center of mass of the system, and assume the motion happens in the $(x,y)$ plane.
The distances of masses $m_1$ and $m_2$ to the center of mass are given by $d_1$ and $d_2$, respectively.
Following the calculation in \cite{Peters1963}, we can obtain the quadrupole moment of the system:
\begin{equation}
\begin{split}
	Q^{xx} &= \mu\, d^2 \cos^2 \psi\,,\\
	Q^{yy} &= \mu\, d^2 \sin^2 \psi\,,\\
	Q^{xy}=Q^{yx} &= \mu\, d^2 \sin \psi \cos \psi\,,
\end{split}
\end{equation}
where we recall that the origin coincides with the center of mass, $\psi$ is the angle that the relative distance vector makes with the $x$ axis, $\mu$ is the reduced mass $m_1 m_2 / (m_1+m_2)$, and $d$ is related to the distances $d_1$ and $d_2$ of masses $m_1$ and $m_2$ to the origin by
\begin{equation}
	d_1 = \frac{m_2}{m_1+m_2}d\;, \;d_2 = \frac{m_1}{m_1+m_2}d\,.
\end{equation}
Considering a general direction $\vec{n}=(\sin \theta \cos \phi, \sin \theta \sin \phi, \cos \theta)$, one obtains
\begin{equation}
n_\alpha n_\beta\, Q^{\alpha\beta} = \mu \, d^2   \sin ^2(\theta ) \cos ^2(\phi -\psi )
\end{equation}
so, using that 
\begin{align}
	d &=\frac{a(1-e^2)}{1+e \cos\psi}\,,\\
	\dot{\psi} &= \frac{\sqrt{G(m_1+m_2)a(1-e^2)}}{d^2}\,,
\end{align}
we obtain 
\begin{equation}
\begin{split}
	&n_\alpha n_\beta\, \dddot{Q}^{\alpha\beta} =- \frac{G^{3/2} m_1 m_2\sqrt{m_1+m_2}}{a^{5/2} \left(1-e^2\right)^{5/2}} \sin ^2(\theta )\\
	&   (e \cos (\psi )+1)^2 \cos (\phi -\psi ) \left(3 e \sin (\phi -2 \psi )\right.\\
	&\left.+5 e \sin (\phi )+8 \sin (\phi -\psi )\right) \,.
\end{split}
\end{equation}
Then Eq.~\eqref{eq:dI} becomes
\begin{equation}
\begin{split}
	\frac{dI}{d\Omega} &= \frac{G^4 m_1^2 m_2^2(m_1+m_2)}{16 \pi  a^5 c^5 \left(1-e^2\right)^5}  \sin ^4(\theta )  (e \cos (\psi )+1)^4\\
	 &\cos^2(\phi -\psi ) (3 e \sin (\phi -2 \psi )+5 e \sin (\phi )+8 \sin (\phi -\psi ))^2\,,
\end{split}
\end{equation}
where $dI/d\Omega$ depends on the radiation direction angles $\theta$ and $\phi$, but also on the orbit angle parameter $\psi$. 
Note that we recover the general relativistic dependence on $G^4/c^5$.
We can average over an orbital period $T$ to get the mean intensity radiated in a specific direction, obtaining
\begin{equation}
\begin{split}
	\left\langle\frac{dI}{d\Omega}\right\rangle_T &= \frac{G^4 m_1^2 m_2^2 (m_1+m_2)}{512 \pi  a^5 c^5 \left(1-e^2\right)^{7/2}}  \sin ^4(\theta ) \\
	&\left(256 + 8e^2 (99 -26\cos(2\phi)) \right.\\
	&\left.+ e^4(102-32\cos(2\phi)-25\cos(4\phi))\right)\,.
\end{split}
\end{equation}
To find the total power radiated in all directions, we just integrate the expression above over the whole sphere, finding
\begin{equation}
\begin{split}
	P = &\frac{16}{15}\frac{G^4 m_1^2 m_2^2 (m_1+m_2)}{a^5 c^5 \left(1-e^2\right)^{7/2}}\\ 
	& \qquad \qquad \times \left(1 + \frac{99}{32} e^2 + \frac{51}{128}e^4\right)\,.
\end{split}
\end{equation}
Though this expression is not exactly the same as general relativity, it is interesting that we find the same order of magnitude ($G^4/c^5$), the same mass dependence on $m_1$ and $m_2$, and a very similar dependence on the eccentricity (quadratic polynomial in $e^2$ in the numerator divided by $(1-e^2)^{7/2}$).

Finally, we can analyze the effect of gravitational waves on matter-energy.
For that, let us consider an observer on Earth, and a gravitational potential given by $\Phi = \Phi_b(\vec{x}) + \Phi_w(\vec{x},t)$, where $\Phi_b$ and $\Phi_w$ represent respectively the contributions from Earth's gravity and from a gravitational wave.
Then, we can get from Eq.~\eqref{eq:partlag} the Euler-Lagrange equations at lowest order:
\begin{equation}
	m_p \frac{dv^i}{dt} = -\partial^i \Phi\,,
\end{equation}
so the gravitational force $f^i_w$ due to the passing wave is simply $f^i_w = -\partial^i \Phi_w$.
This concludes our gravitational wave analysis.

\section{Strong gravity behavior}
\label{sec:strong}

\subsection{Exact solutions for the gravitational potential}

In this section we look for an exact vacuum solution of the theory.
To approach this problem, it is clear from Eq.~\eqref{eq:fieldeq} that we have to specify $f_2$ and $f_3$. 
Our choice will be motivated by simplicity. Since there are no constraints on $f_2$ in the PPN limit (except for $f_2=\Phi/c^2 + \mathcal{O}(c^{-4})$), we consider $f_2=\Phi/c^2$. Also, we keep $ \beta = -2 $.
For $f_3$, there is one more constraint coming from the WEP requirement and the PPN analysis: $\alpha = -\frac{1}{2}$. 
Therefore, we consider the simplest possibility which is $f_3(\Phi) = \Phi/c^2 -\frac{1}{2} \Phi^2/c^4$.
With these assumptions, we can use Eq.~\eqref{eq:f1expr} to write
\begin{equation}\label{eq:f1strong}
	f_1 = \exp\left(\frac{4\Phi}{c^2} - \frac{\Phi^2}{c^4}\right)\,.
\end{equation}
To find a vacuum solution, it is also convenient to use the auxiliary gravitational field $ \tilde{\Phi} $ defined in Eq.~\eqref{eq:phitilde}.
Recall that $ \tilde{\Phi} = c^2 g(\Phi/c^2)$ where $ g' = \sqrt{f_1} $ and $ g(0) = 0 $. 
One can obtain in this case an explicit formula for $ g $:
\begin{equation}
\label{eq:g}
	g(\Phi/c^2) = e^2 \sqrt{\frac{\pi}{2}} \left[\erf\left(\frac{\Phi/c^2-2}{\sqrt{2}}\right) + \erf\left(\sqrt{2}\right)\right]\,.
\end{equation}
Using Eq.~\eqref{eq:fieldeq}, the vacuum field equation for $ \tilde{\Phi} $ becomes simply
\begin{equation}\label{eq:vactilde}
	\square \tilde{\Phi} = 0\,.
\end{equation}
In a static, spherically symmetric situation, one can exactly solve Eq.~\eqref{eq:vactilde} obtaining
\begin{equation}
	\tilde{\Phi}(r) = -\frac{GM}{r}\,.
\end{equation}
Using Eq.~\eqref{eq:g}, we then obtain
\begin{equation}\label{eq:phistrong1}
	\frac{\Phi}{c^2} = \sqrt{2} \erf^{-1}\left(-\sqrt{\frac{2}{\pi}} e^{-2} \frac{m}{r}-\erf(\sqrt{2})\right)+2\,,
\end{equation}
which has a singularity when $ r_s = 2.37 m $.
Notice that $ r_s $ is close to the GR Schwarzschild radius $ r = 2m $, but the two coordinates do not necessarily describe equivalent coordinate systems.

\begin{figure}[h]
	\centering
	\includegraphics[width=\columnwidth]{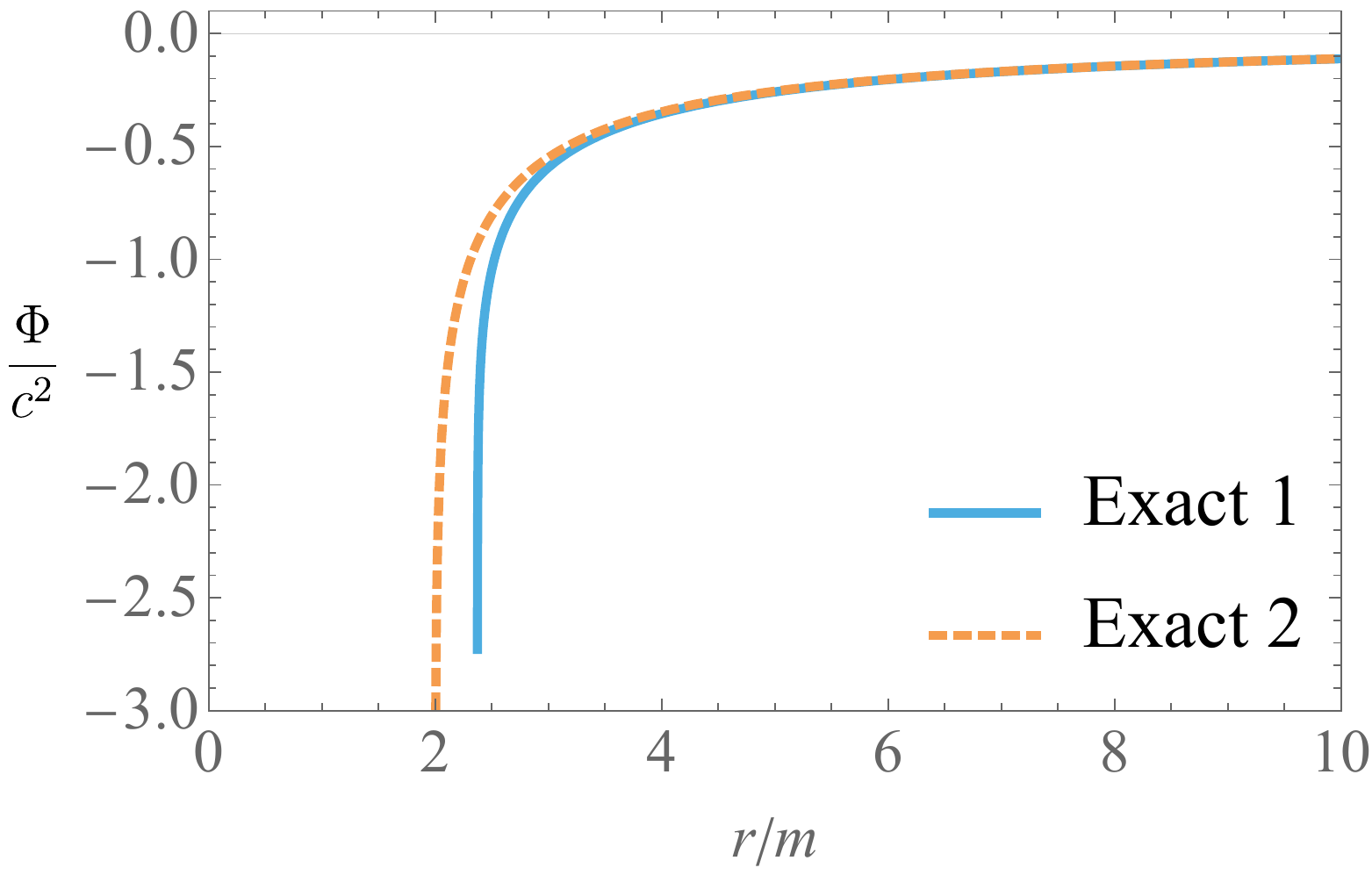}
	\caption{Two gravitational potentials for different $ f_2 $ in the strong field regime. Exact 1 and 2 represent respectively Eqs.~\eqref{eq:phistrong1} and \eqref{eq:phistrong2}.} 
	\label{fig:phistrong}
\end{figure}

Another possible gravitational field could be given by adding a term proportional to $ \Phi^2 $ in $ f_2 $ (which is not constrained in the weak field conditions) such that $ f_1 $ would be more simply given by
\begin{equation}\label{eq:f1strong2}
	f_1 = \exp \left(\frac{4 \Phi}{c^2}\right)\,.
\end{equation}
In this case $ g $ is simpler and given by
\begin{equation}\label{eq:g2}
	g = \frac{\exp(2\Phi/c^2) -1}{2}\,,
\end{equation}
which implies 
\begin{equation}\label{eq:phistrong2}
	\frac{\Phi}{c^2} = \frac{1}{2} \log\left(1-\frac{2m}{r}\right)\,.
\end{equation}
In this case, the singularity happens at exactly $ r_s = 2m $, which matches GR's Schwarzschild radius.
The potentials in Eqs.~\eqref{eq:phistrong1} and \eqref{eq:phistrong2} are shown in Fig.~\ref{fig:phistrong}.
This indicates that the presence of a pole (infinite redshift surface) in $ \Phi $ at around $ r \simeq 2m $ may be a feature of these types of scalar theories of gravity.
A study of strong gravitational fields could thus provide constraints in the expressions of $ f_2 $ and $ f_3 $ at higher order.

Let us now focus on the simpler potential given in Eq.~\eqref{eq:phistrong2} and study the effect it has in particles trajectories.

\subsection{Test particle trajectories}

\begin{figure}[h]
 \centering
    \includegraphics[width=\columnwidth]{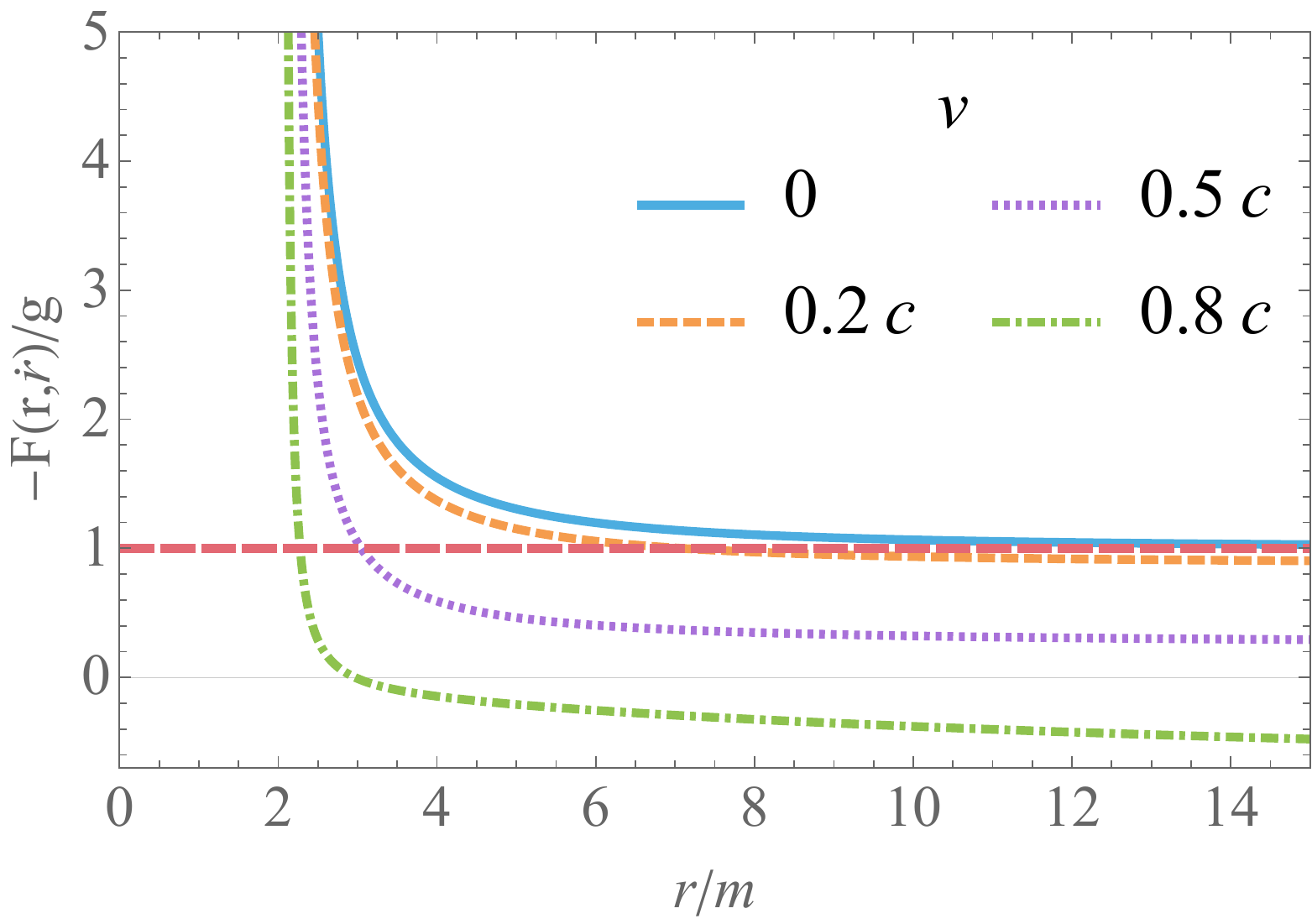}
 \caption{Ratio of $\ddot{r}$ and $ g $ as a function of $r$ for different velocities $v\equiv\dot{r}$. The horizontal line represents Newtonian gravity.} 
\label{fig:forceplot}
\end{figure}

In this section, we calculate the force particles are subject to in the potential given in Eq.~\eqref{eq:phistrong2}. 
We use the exact particle Lagrangian given in Eq.~\eqref{eq:partlag2} to solve for the radial acceleration a particle would have as a function of its radial position and for now we only consider radial motion.
Calculating the Euler-Lagrange equations for the particle Lagrangian and solving for $\ddot{r}$, one gets a rather lengthy expression $\ddot{r}=F(r,\dot{r})$. 
We can then compare with Newtonian gravitation where $ \ddot{r} = - g \equiv -GM/r^2 $.
The ratio $ -F(r,\dot{r})/g $ is shown  in Fig.~\ref{fig:forceplot}.
One sees that the acceleration felt by a particle depends on its speed $ \dot{r} $, and becomes proportional to $ -g $ at large distances. 
The constant of proportionality depends on the speed. Specifically, we have
\begin{equation}\label{eq:Flimit}
	F(r,\dot{r}) = -\frac{GM}{r^2}\left(1 - 3\frac{\dot{r}^2}{c^2}\right) + \mathcal{O}(r^{-3})\,,
\end{equation}  
which matches the weak field limit of GR as well \cite{McGruder1982,Gorkavyi2016}.
So this scalar theory also has the GR property that gravity becomes repulsive for a radial motion when $ \dot{r} > c/\sqrt{3} $ for an observer at infinity.
Moreover, there is a pole of $ F(r,\dot{r}) $ near $ r = 2m $ which is slightly velocity-dependent (because the force depends on the speed).
For concreteness its location $ r_P $ is given by
\begin{equation}\label{eq:forcepole}
	r_P = \frac{2 m}{1-\exp \left(\frac{2 \left(-\sqrt{10 c^2 \dot{r}^2+19 c^4+7 \dot{r}^4}-3 c^2-3 \dot{r}^2\right)}{5 c^2+\dot{r}^2}\right)} \,,
\end{equation}
and we have $ r_P > 2m $. For $ \dot{r} = 0 $, we have $ r_P = 2.11 m $, and for $ \dot{r} = c $, we have  $ r_P = 2.04 m $.
This pole causes a infinite tidal forces and so ``spaghettifies'' any in-falling object, in the same way a black hole would in GR.

\begin{figure}[h]
 \centering
    \includegraphics[width=\columnwidth]{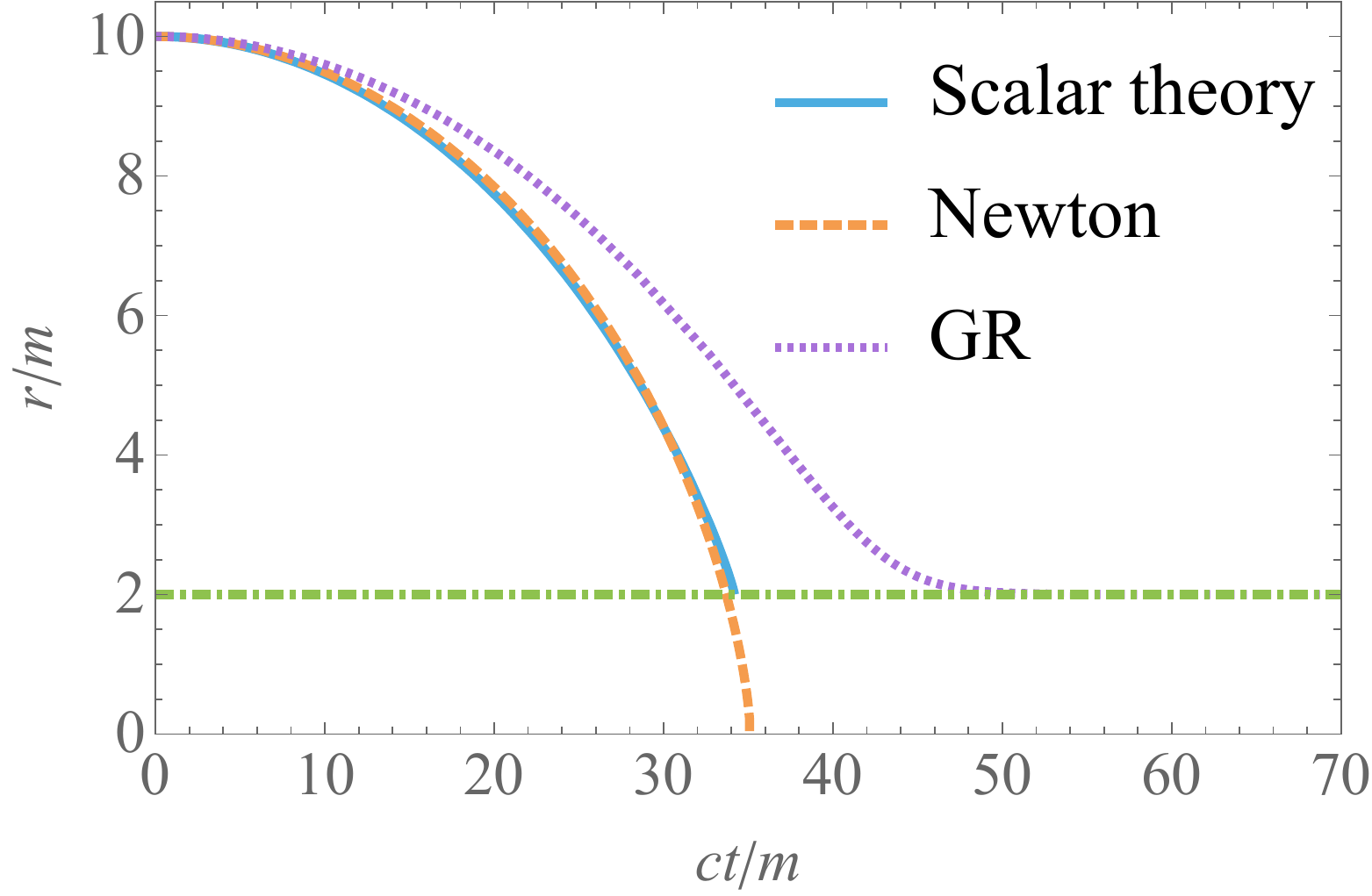}
 \caption{$r$ as a function of $t$ for a free falling particle released at rest from $r_0=10\, m$, with $m=1$, for this scalar field theory, Newtonian gravitation and Einstein's general relativity in Schwarzschild coordinates.} 
\label{fig:freefall}
\end{figure}

Finally, we can compute the trajectory of a free falling particle released at a distance $r_0$ from the origin. 
This is plotted in Fig.~\ref{fig:freefall}, and compared to Newtonian gravity and Einstein's general relativity in Schwarzschild coordinates. 
We notice that, unlike in GR, the trajectory in the scalar theory does not asymptotically flatten out at $ r=2m $, but instead reaches $ r=2m $ in a finite time.
This indicates that observations in the strong field can help differentiate between GR and this scalar theory of gravity.

\section{Conclusion} \label{sec:discussion}

We built a Lorentz invariant, classical scalar theory of gravitation that, unlike Nordstr\"{o}m's theory, is viable in the sense that it predicts the same effects as general relativity in the solar system. 
The main purpose of this work was to show that scalar theories of gravitation could explain all solar system tests.
This theory is self-consistent in that the particle equations of motion and the field equation come from the same Lagrangian density. 
Moreover, gravitational energy gravitates in the same way as any other type of energy.
The conditions derived from the equivalence principle make the theory match general relativity at the post-Newtonian level. 
Any other combination of parameters would not match GR at this order.
In particular, this strongly suggests that this scalar theory is not a weak field limit of a scalar-tensor theory, since the PPN parameters of such theories in general differ from GR~\cite{Scharer2014}.

We calculated the gravitational radiation that this scalar theory predicts for a Hulse-Taylor binary pulsar system, getting results very similar to GR~\cite{Will2014}. 
In particular, we find the same order of magnitude for the energy loss together with quadrupole emission. 
We find also that gravitational waves propagate at the speed of light.
This preliminary calculation can be further refined in future work.
Nevertheless, given the similarity with GR, we can say that this emission of gravitational waves would very likely mimic the profiles observed by LIGO and Virgo for black hole and neutron star mergers~\cite{LIGO2016,LIGOVIRGO2017,LIGOVIRGO2017-NS,LIGOVirgoFermi2017}.

We found effects very similar to GR even in the strong field regime, observing a physical barrier that would ``spaghettify'' any observer at around the Schwarzschild radius. 
A difference from GR happens for a distant observer, which observes the infalling object hitting the ``event horizon'' in a finite time in this theory.
With the black hole observations we have now, it does not seem very straightforward to compare both theories in this case.

Furthermore, cosmology has now validated GR through the great success of the $ \Lambda $CDM model. So another way to confirm or discard this scalar theory would be to study its cosmological predictions.

\section*{ACKNOWLEDGMENTS}

It is a pleasure to thank Rodrigo Vicente and Jos\'{e} P.~S. Lemos for a careful review of this paper which improved its quality, and KIPAC for financial support. 
We also want to thank anonymous referees for very valuable suggestions that greatly improved the quality of this work.

\appendix

\section{Term in the general Lagrangian that would violate special relativity}
\label{app:extraterm}

In this Appendix, we show why $f_2(0)$ must vanish in order for the theory to obey special relativity.

Let us assume $f_2(0) > 0$ in Eq.~\eqref{eq:genlag}. 
In the weak field limit, the corresponding term would then be $\beta T^{ab} \frac{(\partial_a \Phi)(\partial_b \Phi)}{(\partial_c \Phi)(\partial^c \Phi)}$.
Such term would violate special relativity dynamics in the following way.
Suppose we have a static, radial potential $\Phi = \Phi(r)$, then the particle Lagrangian from Eq.~\eqref{eq:partlag} would be 
\begin{equation}
	L_{\rm part} = -m_p c^2 \sqrt{1-\frac{v^2}{c^2}} \left( 1 + \frac{\dot{r}^2}{1-\frac{v^2}{c^2}} + \mathcal{O}(\Phi/c^2)\right)\,,
\end{equation}
where $m_p$ is the mass of the particle and $v$ its coordinate velocity. Assuming the particle has purely radial motion, we find the conserved momentum to be $m \dot{r}/\left(1-\frac{v^2}{c^2}\right)^{3/2}$, which is in clear disagreement with special relativity when the field is very weak. Therefore, we $f_2(0)$ must vanish in Eq.~\eqref{eq:genlag} if we want the theory to satisfy special relativity.

\section{Derivation of effective metric}
\label{app:effectivemetric}

In this short appendix, we derive the form of the effective metric for a static gravitational field at first post-Newtonian order.
From the line element in Eq.~\eqref{eq:effmetric}, we obtain
\begin{equation}
\begin{split}
ds_{\rm eff}^2 &= -f^2 d\tau^2 \\
&= -\left(1+ f_3 - \beta f_2 \frac{(u^a \partial_a \Phi)^2}{w\, c^2}\right)^2 d\tau^2\\
&= -\left(1+ \frac{\Phi}{c^2} + \alpha \frac{\Phi^2}{c^4} - \beta \frac{\Phi}{c^2} \frac{(u^a \partial_a \Phi)^2}{w\, c^2}\right)^2 d\tau^2\\
&= -\left(1+ \frac{2\Phi}{c^2} + (2\alpha+1) \frac{\Phi^2}{c^4} - 2\beta \frac{\Phi}{c^2} \frac{(u^a \partial_a \Phi)^2}{w\, c^2}\right) d\tau^2 \\
&= dx^a dx^b \times\\
&\left[\left(1+ \frac{2\Phi}{c^2} + (2\alpha+1) \frac{\Phi^2}{c^4}\right)\eta_{ab} + 2\beta \frac{\Phi}{c^2} \frac{(\partial_a \Phi)(\partial_b \Phi)}{w}\right]\\
&=g_{ab}dx^a dx^b\,.
\end{split}
\end{equation}
where we used $u^a = dx^a/d\tau$, and we defined $g_{ab}$ as
\begin{multline}
g_{ab} = \left(1+ \frac{2\Phi}{c^2} + (2\alpha+1) \frac{\Phi^2}{c^4}\right)\eta_{ab}\\
 + 2\beta \frac{\Phi}{c^2} \frac{(\partial_a \Phi)(\partial_b \Phi)}{w}\,.
\end{multline} 
Note that, very conveniently, at the first PPN order the effective metric is independent of the 4-velocity of the particle $u^a$. 
If that were the case, then the whole effective metric formalism would be more delicate and it would be harder to compare it with the standard PPN metric.

We can use this effective metric to write in a simpler way the equation of motion for a point particle given in Eq.~\eqref{eq:partEOM}.
At the first post-Newtonian order, the equations of motion become simply
\begin{equation}
	\frac{d^2x^a}{d\lambda^2} + \Gamma^a_{bc} \frac{dx^b}{d\lambda}\frac{dx^c}{d\lambda} = 0\,,
\end{equation}
where $\lambda$ is an affine parameter and $\Gamma^a_{bc}$ are the Christoffel symbols of the metric $g_{ab}$.

\bibliography{ScalarGravity}

\begin{thebibliography}{44}%
\makeatletter
\providecommand \@ifxundefined [1]{%
 \@ifx{#1\undefined}
}%
\providecommand \@ifnum [1]{%
 \ifnum #1\expandafter \@firstoftwo
 \else \expandafter \@secondoftwo
 \fi
}%
\providecommand \@ifx [1]{%
 \ifx #1\expandafter \@firstoftwo
 \else \expandafter \@secondoftwo
 \fi
}%
\providecommand \natexlab [1]{#1}%
\providecommand \enquote  [1]{``#1''}%
\providecommand \bibnamefont  [1]{#1}%
\providecommand \bibfnamefont [1]{#1}%
\providecommand \citenamefont [1]{#1}%
\providecommand \href@noop [0]{\@secondoftwo}%
\providecommand \href [0]{\begingroup \@sanitize@url \@href}%
\providecommand \@href[1]{\@@startlink{#1}\@@href}%
\providecommand \@@href[1]{\endgroup#1\@@endlink}%
\providecommand \@sanitize@url [0]{\catcode `\\12\catcode `\$12\catcode
  `\&12\catcode `\#12\catcode `\^12\catcode `\_12\catcode `\%12\relax}%
\providecommand \@@startlink[1]{}%
\providecommand \@@endlink[0]{}%
\providecommand \url  [0]{\begingroup\@sanitize@url \@url }%
\providecommand \@url [1]{\endgroup\@href {#1}{\urlprefix }}%
\providecommand \urlprefix  [0]{URL }%
\providecommand \Eprint [0]{\href }%
\providecommand \doibase [0]{https://doi.org/}%
\providecommand \selectlanguage [0]{\@gobble}%
\providecommand \bibinfo  [0]{\@secondoftwo}%
\providecommand \bibfield  [0]{\@secondoftwo}%
\providecommand \translation [1]{[#1]}%
\providecommand \BibitemOpen [0]{}%
\providecommand \bibitemStop [0]{}%
\providecommand \bibitemNoStop [0]{.\EOS\space}%
\providecommand \EOS [0]{\spacefactor3000\relax}%
\providecommand \BibitemShut  [1]{\csname bibitem#1\endcsname}%
\let\auto@bib@innerbib\@empty
\bibitem [{\citenamefont {Nordstr{\"{o}}m}(1912)}]{Nordstrom1912}%
  \BibitemOpen
  \bibfield  {author} {\bibinfo {author} {\bibfnamefont {G.}~\bibnamefont
  {Nordstr{\"{o}}m}},\ }\bibfield  {title} {\bibinfo {title}
  {{Relativit{\"{a}}tsprinzip und Gravitation}},\ }\href@noop {} {\bibfield
  {journal} {\bibinfo  {journal} {Physikalische Zeitschrift Boston Studies in
  the Philosophy of Science}\ }\textbf {\bibinfo {volume} {13}},\ \bibinfo
  {pages} {1126} (\bibinfo {year} {1912})}\BibitemShut {NoStop}%
\bibitem [{\citenamefont {Nordstr{\"{o}}m}(1914)}]{Nordstrom1913}%
  \BibitemOpen
  \bibfield  {author} {\bibinfo {author} {\bibfnamefont {G.}~\bibnamefont
  {Nordstr{\"{o}}m}},\ }\bibfield  {title} {\bibinfo {title} {{Die Fallgesetze
  und Planetenbewegungen in der Relativit{\"{a}}tstheorie}},\ }\href@noop {}
  {\bibfield  {journal} {\bibinfo  {journal} {Annalen der Physik}\ }\textbf
  {\bibinfo {volume} {43}},\ \bibinfo {pages} {1101} (\bibinfo {year}
  {1914})}\BibitemShut {NoStop}%
\bibitem [{\citenamefont {Wellner}\ and\ \citenamefont
  {Sandri}(1964)}]{Wellner1964}%
  \BibitemOpen
  \bibfield  {author} {\bibinfo {author} {\bibfnamefont {M.}~\bibnamefont
  {Wellner}}\ and\ \bibinfo {author} {\bibfnamefont {G.}~\bibnamefont
  {Sandri}},\ }\bibfield  {title} {\bibinfo {title} {{Scalar Gravitation}},\
  }\href {https://doi.org/10.1119/1.1970068} {\bibfield  {journal} {\bibinfo
  {journal} {American Journal of Physics}\ }\textbf {\bibinfo {volume} {32}},\
  \bibinfo {pages} {36} (\bibinfo {year} {1964})}\BibitemShut {NoStop}%
\bibitem [{\citenamefont {Norton}(1992)}]{Norton1992}%
  \BibitemOpen
  \bibfield  {author} {\bibinfo {author} {\bibfnamefont {J.~D.}\ \bibnamefont
  {Norton}},\ }\bibfield  {title} {\bibinfo {title} {{Einstein,
  Nordstr$\backslash$"{\{}o{\}}m and the early demise of scalar,
  Lorentz-covariant theories of gravitation}},\ }\href
  {https://doi.org/10.1007/BF00375886} {\bibfield  {journal} {\bibinfo
  {journal} {Archive for History of Exact Sciences}\ }\textbf {\bibinfo
  {volume} {45}},\ \bibinfo {pages} {17} (\bibinfo {year} {1992})}\BibitemShut
  {NoStop}%
\bibitem [{\citenamefont {Einstein}\ and\ \citenamefont
  {Fokker}(1913)}]{Einstein1914}%
  \BibitemOpen
  \bibfield  {author} {\bibinfo {author} {\bibfnamefont {A.}~\bibnamefont
  {Einstein}}\ and\ \bibinfo {author} {\bibfnamefont {A.~D.}\ \bibnamefont
  {Fokker}},\ }\bibfield  {title} {\bibinfo {title} {{Die Nordstr{\"{o}}msche
  Gravitationstheorie vom Standpunkt des absoluten
  Differentialkalk{\"{u}}ls}},\ }\href@noop {} {\bibfield  {journal} {\bibinfo
  {journal} {Annalen der Physik}\ }\textbf {\bibinfo {volume} {349}},\ \bibinfo
  {pages} {321} (\bibinfo {year} {1913})}\BibitemShut {NoStop}%
\bibitem [{\citenamefont {Deruelle}(2011)}]{Deruelle2011}%
  \BibitemOpen
  \bibfield  {author} {\bibinfo {author} {\bibfnamefont {N.}~\bibnamefont
  {Deruelle}},\ }\bibfield  {title} {\bibinfo {title} {{Nordstr{\"{o}}m's
  scalar theory of gravity and the equivalence principle}},\ }\href
  {https://doi.org/10.1007/s10714-011-1247-x} {\bibfield  {journal} {\bibinfo
  {journal} {General Relativity and Gravitation}\ }\textbf {\bibinfo {volume}
  {43}},\ \bibinfo {pages} {3337} (\bibinfo {year} {2011})},\ \Eprint
  {https://arxiv.org/abs/1104.4608v1} {arXiv:1104.4608v1} \BibitemShut
  {NoStop}%
\bibitem [{\citenamefont {Einstein}(1916)}]{Einstein1916}%
  \BibitemOpen
  \bibfield  {author} {\bibinfo {author} {\bibfnamefont {A.}~\bibnamefont
  {Einstein}},\ }\bibfield  {title} {\bibinfo {title} {{Die Grundlage der
  allgemeinen Relativit{\"{a}}tstheorie}},\ }\href@noop {} {\bibfield
  {journal} {\bibinfo  {journal} {Annalen der Physik}\ }\textbf {\bibinfo
  {volume} {354}},\ \bibinfo {pages} {769} (\bibinfo {year}
  {1916})}\BibitemShut {NoStop}%
\bibitem [{\citenamefont {Ni}(2016{\natexlab{a}})}]{Ni2016}%
  \BibitemOpen
  \bibfield  {author} {\bibinfo {author} {\bibfnamefont {W.-T.}\ \bibnamefont
  {Ni}},\ }\bibfield  {title} {\bibinfo {title} {{Solar-system tests of the
  relativistic gravity}},\ }\href {https://doi.org/10.1142/S0218271816300032}
  {\bibfield  {journal} {\bibinfo  {journal} {International Journal of Modern
  Physics D}\ }\textbf {\bibinfo {volume} {25}},\ \bibinfo {pages} {1630003}
  (\bibinfo {year} {2016}{\natexlab{a}})},\ \Eprint
  {https://arxiv.org/abs/1611.06025} {arXiv:1611.06025 [gr-qc]} \BibitemShut
  {NoStop}%
\bibitem [{\citenamefont {Shapiro}\ and\ \citenamefont
  {Teukolsky}(1993)}]{Shapiro1993}%
  \BibitemOpen
  \bibfield  {author} {\bibinfo {author} {\bibfnamefont {S.~L.}\ \bibnamefont
  {Shapiro}}\ and\ \bibinfo {author} {\bibfnamefont {S.~A.}\ \bibnamefont
  {Teukolsky}},\ }\bibfield  {title} {\bibinfo {title} {{Scalar gravitation: A
  laboratory for numerical relativity}},\ }\href
  {https://doi.org/10.1103/PhysRevD.47.1529} {\bibfield  {journal} {\bibinfo
  {journal} {Physical Review D}\ }\textbf {\bibinfo {volume} {47}},\ \bibinfo
  {pages} {1529} (\bibinfo {year} {1993})}\BibitemShut {NoStop}%
\bibitem [{\citenamefont {Watt}\ and\ \citenamefont {Misner}(1999)}]{Watt1999}%
  \BibitemOpen
  \bibfield  {author} {\bibinfo {author} {\bibfnamefont {K.}~\bibnamefont
  {Watt}}\ and\ \bibinfo {author} {\bibfnamefont {C.~W.}\ \bibnamefont
  {Misner}},\ }\bibfield  {title} {\bibinfo {title} {{Relativistic Scalar
  Gravity: A Laboratory for Numerical Relativity}},\ }\href
  {http://arxiv.org/abs/gr-qc/9910032} {\bibfield  {journal} {\bibinfo
  {journal} {arXiv}\ ,\ \bibinfo {pages} {7}} (\bibinfo {year} {1999})},\
  \Eprint {https://arxiv.org/abs/9910032} {arXiv:9910032 [gr-qc]} \BibitemShut
  {NoStop}%
\bibitem [{\citenamefont {Bergmann}(1956)}]{Bergmann1956}%
  \BibitemOpen
  \bibfield  {author} {\bibinfo {author} {\bibfnamefont {O.}~\bibnamefont
  {Bergmann}},\ }\bibfield  {title} {\bibinfo {title} {{Scalar Field Theory as
  a Theory of Gravitation. I}},\ }\href {https://doi.org/10.1119/1.1934129}
  {\bibfield  {journal} {\bibinfo  {journal} {American Journal of Physics}\
  }\textbf {\bibinfo {volume} {24}},\ \bibinfo {pages} {38} (\bibinfo {year}
  {1956})}\BibitemShut {NoStop}%
\bibitem [{\citenamefont {Page}\ and\ \citenamefont
  {Tupper}(1968)}]{Pagetupper1968}%
  \BibitemOpen
  \bibfield  {author} {\bibinfo {author} {\bibfnamefont {C.}~\bibnamefont
  {Page}}\ and\ \bibinfo {author} {\bibfnamefont {B.}~\bibnamefont {Tupper}},\
  }\bibfield  {title} {\bibinfo {title} {{Scalar gravitational theories with
  variable velocity of light}},\ }\href@noop {} {\bibfield  {journal} {\bibinfo
   {journal} {MNRAS}\ }\textbf {\bibinfo {volume} {138}},\ \bibinfo {pages}
  {67} (\bibinfo {year} {1968})}\BibitemShut {NoStop}%
\bibitem [{\citenamefont {Ni}(1973)}]{Ni1973}%
  \BibitemOpen
  \bibfield  {author} {\bibinfo {author} {\bibfnamefont {W.-T.}\ \bibnamefont
  {Ni}},\ }\bibfield  {title} {\bibinfo {title} {{A New Theory of Gravity}},\
  }\href {https://doi.org/10.1103/PhysRevD.7.2880} {\bibfield  {journal}
  {\bibinfo  {journal} {Physical Review D}\ }\textbf {\bibinfo {volume} {7}},\
  \bibinfo {pages} {2880} (\bibinfo {year} {1973})}\BibitemShut {NoStop}%
\bibitem [{\citenamefont {Novello}\ \emph {et~al.}(2013)\citenamefont
  {Novello}, \citenamefont {Bittencourt}, \citenamefont {Moschella},
  \citenamefont {Goulart}, \citenamefont {Salim},\ and\ \citenamefont
  {Toniato}}]{Novello2012}%
  \BibitemOpen
  \bibfield  {author} {\bibinfo {author} {\bibfnamefont {M.}~\bibnamefont
  {Novello}}, \bibinfo {author} {\bibfnamefont {E.}~\bibnamefont
  {Bittencourt}}, \bibinfo {author} {\bibfnamefont {U.}~\bibnamefont
  {Moschella}}, \bibinfo {author} {\bibfnamefont {E.}~\bibnamefont {Goulart}},
  \bibinfo {author} {\bibfnamefont {J.}~\bibnamefont {Salim}},\ and\ \bibinfo
  {author} {\bibfnamefont {J.}~\bibnamefont {Toniato}},\ }\bibfield  {title}
  {\bibinfo {title} {{Geometric scalar theory of gravity}},\ }\href
  {https://doi.org/10.1088/1475-7516/2013/06/014} {\bibfield  {journal}
  {\bibinfo  {journal} {Journal of Cosmology and Astroparticle Physics}\
  }\textbf {\bibinfo {volume} {2013}}\bibfield  {number} {\bibinfo  {number} {
  (06)},\ \bibinfo {pages} {014}},\ }\Eprint {https://arxiv.org/abs/1212.0770}
  {arXiv:1212.0770} \BibitemShut {NoStop}%
\bibitem [{\citenamefont {Ni}(1972)}]{Ni1972}%
  \BibitemOpen
  \bibfield  {author} {\bibinfo {author} {\bibfnamefont {W.-T.}\ \bibnamefont
  {Ni}},\ }\bibfield  {title} {\bibinfo {title} {{Theoretical Frameworks for
  Testing Relativistic Gravity. IV - A Compendium of Metric Theories of Gravity
  and Their POST Newtonian Limits}},\ }\href {https://doi.org/10.1086/151677}
  {\bibfield  {journal} {\bibinfo  {journal} {The Astrophysical Journal}\
  }\textbf {\bibinfo {volume} {176}},\ \bibinfo {pages} {769} (\bibinfo {year}
  {1972})}\BibitemShut {NoStop}%
\bibitem [{\citenamefont {Bragan{\c{c}}a}\ and\ \citenamefont
  {Lemos}(2018)}]{Braganca2018}%
  \BibitemOpen
  \bibfield  {author} {\bibinfo {author} {\bibfnamefont {D.~P.}\ \bibnamefont
  {Bragan{\c{c}}a}}\ and\ \bibinfo {author} {\bibfnamefont {J.~P.}\
  \bibnamefont {Lemos}},\ }\bibfield  {title} {\bibinfo {title} {{Stratified
  scalar field theories of gravitation with self-energy term and effective
  particle Lagrangian}},\ }\href
  {https://doi.org/10.1140/epjc/s10052-018-6006-7} {\bibfield  {journal}
  {\bibinfo  {journal} {European Physical Journal C}\ }\textbf {\bibinfo
  {volume} {78}},\ \bibinfo {pages} {1} (\bibinfo {year} {2018})}\BibitemShut
  {NoStop}%
\bibitem [{\citenamefont {Gupta}(1957)}]{Gupta1957}%
  \BibitemOpen
  \bibfield  {author} {\bibinfo {author} {\bibfnamefont {S.~N.}\ \bibnamefont
  {Gupta}},\ }\bibfield  {title} {\bibinfo {title} {{Einstein's and Other
  Theories of Gravitation}},\ }\href
  {https://doi.org/10.1103/RevModPhys.29.334} {\bibfield  {journal} {\bibinfo
  {journal} {Reviews of Modern Physics}\ }\textbf {\bibinfo {volume} {29}},\
  \bibinfo {pages} {334} (\bibinfo {year} {1957})}\BibitemShut {NoStop}%
\bibitem [{\citenamefont {Harvey}(1965)}]{Harvey1965}%
  \BibitemOpen
  \bibfield  {author} {\bibinfo {author} {\bibfnamefont {A.~L.}\ \bibnamefont
  {Harvey}},\ }\bibfield  {title} {\bibinfo {title} {{Brief Review of
  Lorentz-Covariant Scalar Theories of Gravitation}},\ }\href
  {https://doi.org/10.1119/1.1971681} {\bibfield  {journal} {\bibinfo
  {journal} {American Journal of Physics}\ }\textbf {\bibinfo {volume} {33}},\
  \bibinfo {pages} {449} (\bibinfo {year} {1965})}\BibitemShut {NoStop}%
\bibitem [{\citenamefont {Misner}\ \emph {et~al.}(1973)\citenamefont {Misner},
  \citenamefont {Thorne},\ and\ \citenamefont {Wheeler}}]{gravitation1973}%
  \BibitemOpen
  \bibfield  {author} {\bibinfo {author} {\bibfnamefont {C.~W.}\ \bibnamefont
  {Misner}}, \bibinfo {author} {\bibfnamefont {K.~S.}\ \bibnamefont {Thorne}},\
  and\ \bibinfo {author} {\bibfnamefont {J.~A.}\ \bibnamefont {Wheeler}},\
  }\href@noop {} {\emph {\bibinfo {title} {{Gravitation}}}}\ (\bibinfo
  {publisher} {W.H.{\~{}}Freeman and Co., San Francisco},\ \bibinfo {year}
  {1973})\BibitemShut {NoStop}%
\bibitem [{\citenamefont {Giulini}(2008)}]{Giulini2008}%
  \BibitemOpen
  \bibfield  {author} {\bibinfo {author} {\bibfnamefont {D.}~\bibnamefont
  {Giulini}},\ }\bibfield  {title} {\bibinfo {title} {{What is (not) wrong with
  scalar gravity?}},\ }\href {https://doi.org/10.1016/j.shpsb.2007.09.001}
  {\bibfield  {journal} {\bibinfo  {journal} {Studies in History and Philosophy
  of Science Part B: Studies in History and Philosophy of Modern Physics}\
  }\textbf {\bibinfo {volume} {39}},\ \bibinfo {pages} {154} (\bibinfo {year}
  {2008})},\ \Eprint {https://arxiv.org/abs/0611100} {arXiv:0611100 [gr-qc]}
  \BibitemShut {NoStop}%
\bibitem [{\citenamefont {Will}(1993)}]{Will1993}%
  \BibitemOpen
  \bibfield  {author} {\bibinfo {author} {\bibfnamefont {C.}~\bibnamefont
  {Will}},\ }\href@noop {} {\emph {\bibinfo {title} {{Theory and Experiment in
  Gravitational Physics - Rev. ed.}}}}\ (\bibinfo  {publisher} {Cambridge
  University Press},\ \bibinfo {year} {1993})\BibitemShut {NoStop}%
\bibitem [{\citenamefont {Will}\ and\ \citenamefont {{Nordtvedt,
  Kenneth}}(1972)}]{Will1972}%
  \BibitemOpen
  \bibfield  {author} {\bibinfo {author} {\bibfnamefont {C.~M.}\ \bibnamefont
  {Will}}\ and\ \bibinfo {author} {\bibfnamefont {J.}~\bibnamefont {{Nordtvedt,
  Kenneth}}},\ }\bibfield  {title} {\bibinfo {title} {{Conservation Laws and
  Preferred Frames in Relativistic Gravity. I. Preferred-Frame Theories and an
  Extended PPN Formalism}},\ }\href {https://doi.org/10.1086/151754} {\bibfield
   {journal} {\bibinfo  {journal} {The Astrophysical Journal}\ }\textbf
  {\bibinfo {volume} {177}},\ \bibinfo {pages} {757} (\bibinfo {year}
  {1972})}\BibitemShut {NoStop}%
\bibitem [{\citenamefont {Lee}\ \emph {et~al.}(1976)\citenamefont {Lee},
  \citenamefont {Ni}, \citenamefont {Caves},\ and\ \citenamefont
  {Will}}]{Will1976}%
  \BibitemOpen
  \bibfield  {author} {\bibinfo {author} {\bibfnamefont {D.~L.}\ \bibnamefont
  {Lee}}, \bibinfo {author} {\bibfnamefont {W.-T.}\ \bibnamefont {Ni}},
  \bibinfo {author} {\bibfnamefont {C.~M.}\ \bibnamefont {Caves}},\ and\
  \bibinfo {author} {\bibfnamefont {C.~M.}\ \bibnamefont {Will}},\ }\bibfield
  {title} {\bibinfo {title} {{Theoretical frameworks for testing relativistic
  gravity. V - Post-Newtonian limit of Rosen's theory}},\ }\href
  {https://doi.org/10.1086/154412} {\bibfield  {journal} {\bibinfo  {journal}
  {The Astrophysical Journal}\ }\textbf {\bibinfo {volume} {206}},\ \bibinfo
  {pages} {555} (\bibinfo {year} {1976})}\BibitemShut {NoStop}%
\bibitem [{\citenamefont {Will}(2014)}]{Will2014}%
  \BibitemOpen
  \bibfield  {author} {\bibinfo {author} {\bibfnamefont {C.~M.}\ \bibnamefont
  {Will}},\ }\bibfield  {title} {\bibinfo {title} {{The Confrontation between
  General Relativity and Experiment}},\ }\href
  {https://doi.org/10.12942/lrr-2014-4} {\bibfield  {journal} {\bibinfo
  {journal} {Living Reviews in Relativity}\ }\textbf {\bibinfo {volume} {17}},\
  \bibinfo {pages} {4} (\bibinfo {year} {2014})},\ \Eprint
  {https://arxiv.org/abs/1403.7377} {arXiv:1403.7377} \BibitemShut {NoStop}%
\bibitem [{\citenamefont {Hohmann}\ \emph {et~al.}(2013)\citenamefont
  {Hohmann}, \citenamefont {J{\"{a}}rv}, \citenamefont {Kuusk},\ and\
  \citenamefont {Randla}}]{Hohmann2013}%
  \BibitemOpen
  \bibfield  {author} {\bibinfo {author} {\bibfnamefont {M.}~\bibnamefont
  {Hohmann}}, \bibinfo {author} {\bibfnamefont {L.}~\bibnamefont {J{\"{a}}rv}},
  \bibinfo {author} {\bibfnamefont {P.}~\bibnamefont {Kuusk}},\ and\ \bibinfo
  {author} {\bibfnamefont {E.}~\bibnamefont {Randla}},\ }\bibfield  {title}
  {\bibinfo {title} {{Post-Newtonian parameters $\gamma$ and $\beta$ of
  scalar-tensor gravity with a general potential}},\ }\href
  {https://doi.org/10.1103/PhysRevD.88.084054} {\bibfield  {journal} {\bibinfo
  {journal} {Physical Review D}\ }\textbf {\bibinfo {volume} {88}},\ \bibinfo
  {pages} {084054} (\bibinfo {year} {2013})},\ \Eprint
  {https://arxiv.org/abs/1309.0031} {arXiv:1309.0031} \BibitemShut {NoStop}%
\bibitem [{\citenamefont {Hohmann}\ \emph {et~al.}(2016)\citenamefont
  {Hohmann}, \citenamefont {J{\"{a}}rv}, \citenamefont {Kuusk}, \citenamefont
  {Randla},\ and\ \citenamefont {Vilson}}]{Hohmann2016}%
  \BibitemOpen
  \bibfield  {author} {\bibinfo {author} {\bibfnamefont {M.}~\bibnamefont
  {Hohmann}}, \bibinfo {author} {\bibfnamefont {L.}~\bibnamefont {J{\"{a}}rv}},
  \bibinfo {author} {\bibfnamefont {P.}~\bibnamefont {Kuusk}}, \bibinfo
  {author} {\bibfnamefont {E.}~\bibnamefont {Randla}},\ and\ \bibinfo {author}
  {\bibfnamefont {O.}~\bibnamefont {Vilson}},\ }\bibfield  {title} {\bibinfo
  {title} {{Post-Newtonian parameter $\gamma$ for multiscalar-tensor gravity
  with a general potential}},\ }\href
  {https://doi.org/10.1103/PhysRevD.94.124015} {\bibfield  {journal} {\bibinfo
  {journal} {Physical Review D}\ }\textbf {\bibinfo {volume} {94}},\ \bibinfo
  {pages} {124015} (\bibinfo {year} {2016})},\ \Eprint
  {https://arxiv.org/abs/1607.02356} {arXiv:1607.02356} \BibitemShut {NoStop}%
\bibitem [{\citenamefont {Hohmann}\ and\ \citenamefont
  {Sch{\"{a}}rer}(2017)}]{Hohmann2017}%
  \BibitemOpen
  \bibfield  {author} {\bibinfo {author} {\bibfnamefont {M.}~\bibnamefont
  {Hohmann}}\ and\ \bibinfo {author} {\bibfnamefont {A.}~\bibnamefont
  {Sch{\"{a}}rer}},\ }\bibfield  {title} {\bibinfo {title} {{Post-Newtonian
  parameters $\gamma$ and $\beta$ of scalar-tensor gravity for a homogeneous
  gravitating sphere}},\ }\href {https://doi.org/10.1103/PhysRevD.96.104026}
  {\bibfield  {journal} {\bibinfo  {journal} {Physical Review D}\ }\textbf
  {\bibinfo {volume} {96}},\ \bibinfo {pages} {104026} (\bibinfo {year}
  {2017})},\ \Eprint {https://arxiv.org/abs/1708.07851} {arXiv:1708.07851}
  \BibitemShut {NoStop}%
\bibitem [{\citenamefont {Ni}(2016{\natexlab{b}})}]{Ni2016-2}%
  \BibitemOpen
  \bibfield  {author} {\bibinfo {author} {\bibfnamefont {W.-T.}\ \bibnamefont
  {Ni}},\ }\bibfield  {title} {\bibinfo {title} {{A nonmetric theory of
  gravity}},\ }\href {https://doi.org/10.1142/S0218271816400174} {\bibfield
  {journal} {\bibinfo  {journal} {International Journal of Modern Physics D}\
  }\textbf {\bibinfo {volume} {25}},\ \bibinfo {pages} {1640017} (\bibinfo
  {year} {2016}{\natexlab{b}})}\BibitemShut {NoStop}%
\bibitem [{\citenamefont {Sanghai}\ and\ \citenamefont
  {Clifton}(2017)}]{Sanghai2017}%
  \BibitemOpen
  \bibfield  {author} {\bibinfo {author} {\bibfnamefont {V.~A.~A.}\
  \bibnamefont {Sanghai}}\ and\ \bibinfo {author} {\bibfnamefont
  {T.}~\bibnamefont {Clifton}},\ }\bibfield  {title} {\bibinfo {title}
  {{Parameterized post-Newtonian cosmology}},\ }\href
  {https://doi.org/10.1088/1361-6382/aa5d75} {\bibfield  {journal} {\bibinfo
  {journal} {Classical and Quantum Gravity}\ }\textbf {\bibinfo {volume}
  {34}},\ \bibinfo {pages} {065003} (\bibinfo {year} {2017})},\ \Eprint
  {https://arxiv.org/abs/1610.08039} {arXiv:1610.08039} \BibitemShut {NoStop}%
\bibitem [{\citenamefont {Giulini}(1997)}]{Giulini1996}%
  \BibitemOpen
  \bibfield  {author} {\bibinfo {author} {\bibfnamefont {D.}~\bibnamefont
  {Giulini}},\ }\bibfield  {title} {\bibinfo {title} {{Consistently
  implementing the field self-energy in Newtonian gravity}},\ }\href
  {https://doi.org/10.1016/S0375-9601(97)00369-1} {\bibfield  {journal}
  {\bibinfo  {journal} {Physics Letters A}\ }\textbf {\bibinfo {volume}
  {232}},\ \bibinfo {pages} {165} (\bibinfo {year} {1997})},\ \Eprint
  {https://arxiv.org/abs/9605011} {arXiv:9605011 [gr-qc]} \BibitemShut
  {NoStop}%
\bibitem [{\citenamefont {Franklin}(2015)}]{Franklin2014}%
  \BibitemOpen
  \bibfield  {author} {\bibinfo {author} {\bibfnamefont {J.}~\bibnamefont
  {Franklin}},\ }\bibfield  {title} {\bibinfo {title} {{Self-consistent,
  self-coupled scalar gravity}},\ }\href {https://doi.org/10.1119/1.4898585}
  {\bibfield  {journal} {\bibinfo  {journal} {American Journal of Physics}\
  }\textbf {\bibinfo {volume} {83}},\ \bibinfo {pages} {332} (\bibinfo {year}
  {2015})},\ \Eprint {https://arxiv.org/abs/gr-qc/1408.3594}
  {arXiv:gr-qc/1408.3594 [gr-qc]} \BibitemShut {NoStop}%
\bibitem [{\citenamefont {Franklin}\ \emph {et~al.}(2016)\citenamefont
  {Franklin}, \citenamefont {Guo}, \citenamefont {Newton},\ and\ \citenamefont
  {Schlosshauer}}]{Franklin2016}%
  \BibitemOpen
  \bibfield  {author} {\bibinfo {author} {\bibfnamefont {J.}~\bibnamefont
  {Franklin}}, \bibinfo {author} {\bibfnamefont {Y.}~\bibnamefont {Guo}},
  \bibinfo {author} {\bibfnamefont {K.~C.}\ \bibnamefont {Newton}},\ and\
  \bibinfo {author} {\bibfnamefont {M.}~\bibnamefont {Schlosshauer}},\
  }\bibfield  {title} {\bibinfo {title} {{The dynamics of the
  Schr{\"{o}}dinger–Newton system with self-field coupling}},\ }\href
  {https://doi.org/10.1088/0264-9381/33/7/075002} {\bibfield  {journal}
  {\bibinfo  {journal} {Classical and Quantum Gravity}\ }\textbf {\bibinfo
  {volume} {33}},\ \bibinfo {pages} {075002} (\bibinfo {year} {2016})},\
  \Eprint {https://arxiv.org/abs/1603.03380} {arXiv:1603.03380} \BibitemShut
  {NoStop}%
\bibitem [{\citenamefont {Weinberg}(1972)}]{WeinbergGravitation1972}%
  \BibitemOpen
  \bibfield  {author} {\bibinfo {author} {\bibfnamefont {S.}~\bibnamefont
  {Weinberg}},\ }\href@noop {} {\emph {\bibinfo {title} {{Gravitation and
  Cosmology: Principles and Applications of the General Theory of
  Relativity}}}}\ (\bibinfo  {publisher} {Wiley},\ \bibinfo {address} {New
  York, NY},\ \bibinfo {year} {1972})\BibitemShut {NoStop}%
\bibitem [{\citenamefont {Zych}\ \emph {et~al.}(2019)\citenamefont {Zych},
  \citenamefont {Rudnicki},\ and\ \citenamefont {Pikovski}}]{Zych2019}%
  \BibitemOpen
  \bibfield  {author} {\bibinfo {author} {\bibfnamefont {M.}~\bibnamefont
  {Zych}}, \bibinfo {author} {\bibfnamefont {{\L}.}~\bibnamefont {Rudnicki}},\
  and\ \bibinfo {author} {\bibfnamefont {I.}~\bibnamefont {Pikovski}},\
  }\bibfield  {title} {\bibinfo {title} {{Gravitational mass of composite
  systems}},\ }\href {https://doi.org/10.1103/PhysRevD.99.104029} {\bibfield
  {journal} {\bibinfo  {journal} {Physical Review D}\ }\textbf {\bibinfo
  {volume} {99}},\ \bibinfo {pages} {104029} (\bibinfo {year} {2019})},\
  \Eprint {https://arxiv.org/abs/1808.05831} {arXiv:1808.05831} \BibitemShut
  {NoStop}%
\bibitem [{\citenamefont {McGruder}(1982)}]{McGruder1982}%
  \BibitemOpen
  \bibfield  {author} {\bibinfo {author} {\bibfnamefont {C.~H.}\ \bibnamefont
  {McGruder}},\ }\bibfield  {title} {\bibinfo {title} {{Gravitational repulsion
  in the Schwarzschild field}},\ }\href
  {https://doi.org/10.1103/PhysRevD.25.3191} {\bibfield  {journal} {\bibinfo
  {journal} {Physical Review D}\ }\textbf {\bibinfo {volume} {25}},\ \bibinfo
  {pages} {3191} (\bibinfo {year} {1982})}\BibitemShut {NoStop}%
\bibitem [{\citenamefont {Weisberg}\ and\ \citenamefont
  {Taylor}(2004)}]{Weisberg2004}%
  \BibitemOpen
  \bibfield  {author} {\bibinfo {author} {\bibfnamefont {J.~M.}\ \bibnamefont
  {Weisberg}}\ and\ \bibinfo {author} {\bibfnamefont {J.~H.}\ \bibnamefont
  {Taylor}},\ }\bibfield  {title} {\bibinfo {title} {{Relativistic Binary
  Pulsar B1913+16: Thirty Years of Observations and Analysis}},\ }\href
  {http://arxiv.org/abs/astro-ph/0407149
  https://arxiv.org/abs/astro-ph/0407149v1} {\bibfield  {journal} {\bibinfo
  {journal} {Binary Radio Pulsars ASP Conference Series}\ } (\bibinfo {year}
  {2004})},\ \Eprint {https://arxiv.org/abs/0407149} {arXiv:0407149 [astro-ph]}
  \BibitemShut {NoStop}%
\bibitem [{\citenamefont {Landau}\ and\ \citenamefont
  {Lifshitz}(1975)}]{Landau1975}%
  \BibitemOpen
  \bibfield  {author} {\bibinfo {author} {\bibfnamefont {L.}~\bibnamefont
  {Landau}}\ and\ \bibinfo {author} {\bibfnamefont {E.}~\bibnamefont
  {Lifshitz}},\ }\href@noop {} {\emph {\bibinfo {title} {{The Classical Theory
  of Fields}}}}\ (\bibinfo  {publisher} {Pergamon Press},\ \bibinfo {address}
  {Oxford},\ \bibinfo {year} {1975})\BibitemShut {NoStop}%
\bibitem [{\citenamefont {Peters}\ and\ \citenamefont
  {Mathews}(1963)}]{Peters1963}%
  \BibitemOpen
  \bibfield  {author} {\bibinfo {author} {\bibfnamefont {P.~C.}\ \bibnamefont
  {Peters}}\ and\ \bibinfo {author} {\bibfnamefont {J.}~\bibnamefont
  {Mathews}},\ }\bibfield  {title} {\bibinfo {title} {{Gravitational Radiation
  from Point Masses in a Keplerian Orbit}},\ }\href
  {https://doi.org/10.1103/PhysRev.131.435} {\bibfield  {journal} {\bibinfo
  {journal} {Physical Review}\ }\textbf {\bibinfo {volume} {131}},\ \bibinfo
  {pages} {435} (\bibinfo {year} {1963})}\BibitemShut {NoStop}%
\bibitem [{\citenamefont {Gorkavyi}\ and\ \citenamefont
  {Vasilkov}(2016)}]{Gorkavyi2016}%
  \BibitemOpen
  \bibfield  {author} {\bibinfo {author} {\bibfnamefont {N.}~\bibnamefont
  {Gorkavyi}}\ and\ \bibinfo {author} {\bibfnamefont {A.}~\bibnamefont
  {Vasilkov}},\ }\bibfield  {title} {\bibinfo {title} {{A repulsive force in
  the Einstein theory}},\ }\href {https://doi.org/10.1093/mnras/stw1517}
  {\bibfield  {journal} {\bibinfo  {journal} {Monthly Notices of the Royal
  Astronomical Society}\ }\textbf {\bibinfo {volume} {461}},\ \bibinfo {pages}
  {2929} (\bibinfo {year} {2016})}\BibitemShut {NoStop}%
\bibitem [{\citenamefont {Sch{\"{a}}rer}\ \emph {et~al.}(2014)\citenamefont
  {Sch{\"{a}}rer}, \citenamefont {Ang{\'{e}}lil}, \citenamefont {Bondarescu},
  \citenamefont {Jetzer},\ and\ \citenamefont {Lundgren}}]{Scharer2014}%
  \BibitemOpen
  \bibfield  {author} {\bibinfo {author} {\bibfnamefont {A.}~\bibnamefont
  {Sch{\"{a}}rer}}, \bibinfo {author} {\bibfnamefont {R.}~\bibnamefont
  {Ang{\'{e}}lil}}, \bibinfo {author} {\bibfnamefont {R.}~\bibnamefont
  {Bondarescu}}, \bibinfo {author} {\bibfnamefont {P.}~\bibnamefont {Jetzer}},\
  and\ \bibinfo {author} {\bibfnamefont {A.}~\bibnamefont {Lundgren}},\
  }\bibfield  {title} {\bibinfo {title} {{Testing scalar-tensor theories and
  PPN parameters in Earth orbit}}\ }\href
  {https://doi.org/10.1103/PhysRevD.90.123005} {10.1103/PhysRevD.90.123005}
  (\bibinfo {year} {2014}),\ \Eprint {https://arxiv.org/abs/1410.7914}
  {arXiv:1410.7914} \BibitemShut {NoStop}%
\bibitem [{\citenamefont {{The LIGO Scientific Collaboration}}\ and\
  \citenamefont {{the Virgo Collaboration}}(2016)}]{LIGO2016}%
  \BibitemOpen
  \bibfield  {author} {\bibinfo {author} {\bibnamefont {{The LIGO Scientific
  Collaboration}}}\ and\ \bibinfo {author} {\bibnamefont {{the Virgo
  Collaboration}}},\ }\bibfield  {title} {\bibinfo {title} {{Observation of
  Gravitational Waves from a Binary Black Hole Merger}},\ }\href
  {https://doi.org/10.1103/PhysRevLett.116.061102} {\bibfield  {journal}
  {\bibinfo  {journal} {Physical Review Letters}\ }\textbf {\bibinfo {volume}
  {116}},\ \bibinfo {pages} {061102} (\bibinfo {year} {2016})},\ \Eprint
  {https://arxiv.org/abs/1602.03837} {arXiv:1602.03837} \BibitemShut {NoStop}%
\bibitem [{\citenamefont {{The LIGO Scientific Collaboration}}\ and\
  \citenamefont {{The Virgo
  Collaboration}}(2017{\natexlab{a}})}]{LIGOVIRGO2017}%
  \BibitemOpen
  \bibfield  {author} {\bibinfo {author} {\bibnamefont {{The LIGO Scientific
  Collaboration}}}\ and\ \bibinfo {author} {\bibnamefont {{The Virgo
  Collaboration}}},\ }\bibfield  {title} {\bibinfo {title} {{GW170814: A
  Three-Detector Observation of Gravitational Waves from a Binary Black Hole
  Coalescence}},\ }\href {https://doi.org/10.1103/PhysRevLett.119.141101}
  {\bibfield  {journal} {\bibinfo  {journal} {Physical Review Letters}\
  }\textbf {\bibinfo {volume} {119}},\ \bibinfo {pages} {141101} (\bibinfo
  {year} {2017}{\natexlab{a}})},\ \Eprint {https://arxiv.org/abs/1709.09660}
  {arXiv:1709.09660} \BibitemShut {NoStop}%
\bibitem [{\citenamefont {{The LIGO Scientific Collaboration}}\ and\
  \citenamefont {{The Virgo
  Collaboration}}(2017{\natexlab{b}})}]{LIGOVIRGO2017-NS}%
  \BibitemOpen
  \bibfield  {author} {\bibinfo {author} {\bibnamefont {{The LIGO Scientific
  Collaboration}}}\ and\ \bibinfo {author} {\bibnamefont {{The Virgo
  Collaboration}}},\ }\bibfield  {title} {\bibinfo {title} {{GW170817:
  Observation of Gravitational Waves from a Binary Neutron Star Inspiral}},\
  }\href {https://doi.org/10.1103/PhysRevLett.119.161101} {\bibfield  {journal}
  {\bibinfo  {journal} {Physical Review Letters}\ }\textbf {\bibinfo {volume}
  {119}},\ \bibinfo {pages} {161101} (\bibinfo {year} {2017}{\natexlab{b}})},\
  \Eprint {https://arxiv.org/abs/1710.05832} {arXiv:1710.05832} \BibitemShut
  {NoStop}%
\bibitem [{\citenamefont {{LIGO Scientific Collaboration}}\ \emph
  {et~al.}(2017)\citenamefont {{LIGO Scientific Collaboration}}, \citenamefont
  {{Virgo Collaboration}}, \citenamefont {Monitor},\ and\ \citenamefont
  {INTEGRAL}}]{LIGOVirgoFermi2017}%
  \BibitemOpen
  \bibfield  {author} {\bibinfo {author} {\bibnamefont {{LIGO Scientific
  Collaboration}}}, \bibinfo {author} {\bibnamefont {{Virgo Collaboration}}},
  \bibinfo {author} {\bibfnamefont {F.~G.-R.~B.}\ \bibnamefont {Monitor}},\
  and\ \bibinfo {author} {\bibnamefont {INTEGRAL}},\ }\bibfield  {title}
  {\bibinfo {title} {{Gravitational Waves and Gamma-rays from a Binary Neutron
  Star Merger: GW170817 and GRB 170817A}},\ }\href
  {https://doi.org/10.3847/2041-8213/aa920c} {\bibfield  {journal} {\bibinfo
  {journal} {The Astrophysical Journal}\ }\textbf {\bibinfo {volume} {848}},\
  \bibinfo {pages} {L13} (\bibinfo {year} {2017})},\ \Eprint
  {https://arxiv.org/abs/1710.05834} {arXiv:1710.05834} \BibitemShut {NoStop}%
\end{thebibliography}%

\end{document}